\documentclass[twocolumn,           
               showpacs,            
               preprintnumbers,     
               aps,                 
               prd,                 
               a4paper,             
               superscriptaddress,  
               nofootinbib,         
               tightenlines,        
               floats,floatfix      
               ]{revtex4}
\usepackage{graphicx}
\usepackage{latexsym}
\usepackage{amsmath,amssymb}        
\usepackage[draft=false]{hyperref}

\def\be{\begin{equation}}
\def\ee{\end{equation}}
\def\bea{\begin{eqnarray}}
\def\eea{\end{eqnarray}}

\begin{document}

 \typeout{Prints "DRAFT" on each page; does not show in TeXView}
 \title{Directional statistics for realistic WIMP direct detection experiments}
 \author{Ben Morgan}
\affiliation{Department of Physics and
 Astronomy, University of Sheffield, Hicks Building, Hounsfield Road,
 Sheffield, S3 7RH, United Kingdom} 
 \author{Anne M. Green} 
\affiliation{Department of Physics and
 Astronomy, University of Sheffield, Hicks Building, Hounsfield Road,
 Sheffield, S3 7RH, United Kingdom  (present address)\\ 
and Astronomy Centre, University
 of Sussex, Brighton BN1 9QH, United Kingdom\\ and Physics Department,
 Stockholm University, Stockholm, S106 91, Sweden}
 \author{Neil J. C. Spooner}
\affiliation{Department of Physics and
 Astronomy, University of Sheffield, Hicks Building, Hounsfield Road,
 Sheffield, S3 7RH, United Kingdom}

\date{\today} 
\pacs{95.35.+d\hfill astro-ph/0408047}
\preprint{astro-ph/0408047}
\begin{abstract}
The direction dependence of the event rate in WIMP direct detection
experiments provides a powerful tool for distinguishing
WIMP events from potential backgrounds.  We use a variety of
(non-parametric) statistical tests to examine the number of events
required to distinguish a WIMP signal from an isotropic background
when the uncertainty in the reconstruction of the nuclear recoil
direction is included in the calculation of the expected signal.  We
consider a range of models for the Milky Way halo, and also study rotational 
symmetry tests aimed at detecting non-sphericity/isotropy of the Milky
Way halo. Finally we examine ways of detecting tidal streams of WIMPs. We find
that if the senses of the recoils are known then of order ten events 
will be sufficient to distinguish a WIMP signal from an isotropic background 
for all of the halo models considered, with the uncertainties in 
reconstructing the recoil direction only mildly increasing the required 
number of events. If the senses of the recoils are not known the number of 
events required is an order of magnitude larger, with a large variation 
between halo models, and the recoil resolution is now an important factor.
The rotational symmetry tests require of order a thousand events to distinguish
between spherical and significantly triaxial halos, however a deviation of the
peak recoil direction from the direction of the solar motion due 
to a tidal stream could be detected with of order 
a hundred events, regardless of whether the sense of the recoils is known.

\end{abstract}
\maketitle

\section{Introduction}
\label{intro}

Weakly Interacting Massive Particle (WIMP) direct
detection experiments aim to directly detect non-baryonic dark matter
via the elastic scattering of WIMPs on detector nuclei, and are presently
reaching the sensitivity required to detect neutralinos (the lightest
supersymmetric particle and an excellent WIMP candidate). Since the
expected event rates are very small ( ${\cal O} (10^{-5} - 1)$ counts
${\rm kg^{-1} day^{-1}}$) distinguishing a putative Weakly Interacting
Massive Particle (WIMP) signal from backgrounds due to, for instance,
neutrons from cosmic-ray induced muons or natural radioactivity, is
crucial.  The Earth's motion about the Sun provides two potential WIMP
smoking guns: i) an annual modulation~\cite{amtheory} and ii) a strong
direction dependence~\cite{dirndep} of the event rate. In principle
the dependence of the differential event rate on the atomic mass of
the detector (see e.g. Refs.~\cite{jkg,ls}) is a third possibility,
however this would require good control of systematics for detectors
composed of different materials.

The DAMA ({\bf DA}rk {\bf MA}tter) collaboration have, using $\sim 100
\, {\rm kg}$ of NaI crystals and an exposure of $\sim 1.1 \times \,
10^{5} \, {\rm kg \, day}$, observed an annual modulation in their
event rate~\cite{dama}, which they interpret as a WIMP
signal. Taken at face value the allowed region of WIMP mass and
cross-section parameter space is incompatible with the exclusion
limits from other experiments (such as the {\bf C}ryogenic {\bf D}ark 
{\bf M}atter {\bf S}earch,
Edelweiss and Zeplin I)~\cite{otherexpt}, however the form of the WIMP
annual modulation signal, in particular its phase, depends sensitively
on the local WIMP velocity
distribution~\cite{amhalo,amevans,amhalog1,amhalog2}~\footnote{Other
theoretical possibilities which could perhaps resolve this conflict
include spin-dependent interactions~\cite{spindep} and inelastic
scattering~\cite{inelas}.}.  It is, however, difficult to exclude the
possibility that the annual modulation could be due to some previously
unidentified background.

The direction dependence of the WIMP scattering rate has several
advantages over the annual modulation. Firstly, the amplitude of the
annual modulation is typically of order a few per-cent~\cite{amtheory,ls}
while the event rate in the forward direction is roughly an order of
magnitude larger than that in the backward
direction~\cite{dirndep,ls}. Secondly it is difficult for the
directional signal to be mimicked by backgrounds; in most cases
(a point source in the lab is a possible exception)
a background which
is anisotropic in the laboratory frame will be isotropic in the
Galactic rest frame as the, time dependent, conversion between the lab
and Galactic co-ordinate frames will wash out any lab specific
features.  Furthermore the mean direction, in the lab, of WIMP induced
recoils will vary over a sidereal day due to the rotation of the
Earth~\cite{dirrot}.  Designing a detector capable of measuring the
directions of sub-100 keV nuclear recoils is however a considerable
challenge. Prototype directional detectors based on roton anisotropy
in liquid He~\cite{lhedet}, phonon anisotropy in BaF
crystals~\cite{bafdet} and the
anisotropic scintillation properties of stilbene~\cite{stildet} have
been tested,  however low pressure gas time projection chambers
(TPCs), such as DRIFT ({\bf D}irectional {\bf R}ecoil {\bf
I}dentification {\bf F}rom {\bf T}racks)~\cite{drift,sean:drift} 
and NEWAGE~\cite{newage}, seem
to offer the best prospects for a workable detector. In discriminating
a WIMP signal from isotropic backgrounds it is therefore crucial to
take into account the accuracy with which recoil directions can be
measured by such a detector.  Furthermore, the direction dependence of
the recoil rate depends on the local WIMP velocity
distribution~\cite{copi:krauss}. Whilst this opens up the exciting
possibility of experimentally probing the local dark matter
distribution if WIMPs are detected (i.e. doing WIMP astronomy), it is
also crucial that uncertainties in this distribution are accounted for
when designing tests for discriminating a WIMP signal from isotropic
backgrounds.

Earlier studies have found that as few as 30 events would be required
to distinguish a WIMP induced signal from isotropic
backgrounds~\cite{copi:krauss,lehner:dir} and that with of order
hundreds of events it may be possible to distinguish between different
models for the Galactic halo~\cite{copi:krauss}. In this paper we
improve on these works by including the uncertainty in the
reconstruction of the nuclear recoil direction and applying
non-parametric tests to unbinned data. In Sec.~\ref{recoil} we
describe our calculation of the nuclear recoil spectrum, including the
modeling of the Milky Way halo (Sec.~\ref{halo}) and the
reconstruction of the recoil direction (Sec.~\ref{detector}).  In
Sec.~\ref{stat} we then apply an array of statistical tests aimed at
probing the isotropy (Sec.~\ref{iso}), rotational symmetry
(Sec.~\ref{rotsymtest}) and mean direction (Sec.~\ref{dirtest}) of a
putative WIMP directional signal, before concluding with discussion of
our results in Sec.~\ref{discuss}. In Appendices A, B and C we outline
the calculation of the Earth's orbital velocity, the statistics used and the
hypothesis testing formalism respectively.

\section{Calculating the nuclear recoil spectrum}
\label{recoil}

\subsection{Modeling the Milky Way halo}
\label{halo}

The simplest possible model of the Milky Way (MW) halo is an
isotropic sphere with density distribution $\rho \propto r^{-2}$, in
which case the velocity distribution at all positions within the halo
is maxwellian: 
\begin{equation} 
\label{fmax}
f_{0}(\vec{v}) = \frac{1}{(2\pi/3)^{3/2}\sigma_0^3}
     \exp{\left(\frac{3|\vec{v}|^2}{2\sigma_0^2}\right)}
\,, 
\end{equation} 
where $\sigma_0$, the velocity dispersion, is
related to the asymptotic circular velocity (which we take throughout
to be $v_{{\rm c}} = 220 \, {\rm km \, s}^{-1}$) by $\sigma_0 =
\sqrt{3/2}\,  v_{{\rm c}}$.

Observations and numerical simulations indicate that galaxy halos are
in fact triaxial and anisotropic and contain substructure.  Numerical
simulations produce both prolate ($c/b>b/a$) and oblate ($c/b<b/a$)
halos (where $(a,b,c)$ are the lengths of the long, intermediate and
short principal axes of the density distribution
respectively)~\cite{shape,js}. To date only cluster sized halos have
been simulated in large numbers.  Ref.~\cite{js} found that the shape
parameters of such large halos lie in the ranges $c/a \sim 0.4-0.6$
and $b/a\sim0.7-0.8$. However the axial ratios vary with radius within
a given halo, depending at least partly on the merger history of the
halo. Ref.~\cite{moore:draco} studied two high resolution simulations
of MW size halos, one halo had $c/a \sim b/a \sim0.4$, the other
$c/a\sim0.5$ and $ b/a \sim0.8$. These simulations contain only
collision-less matter, and the addition of baryons leads to halos that
are less triaxial, especially in the central regions~\cite{sph}.
These ranges of axial ratios are in rough agreement with observational
measurements of the shape of dark matter halos (for reviews see
e.g. Ref.~\cite{shapeobs}).

The two MW-like halos in Ref.~\cite{moore:draco} have velocity anisotropy  
at the solar radius ($r=R_{0} \approx 8 \, {\rm kpc}$) 
$\beta(R_{0})\sim0.3\pm0.1$ 
where $\beta$ is defined as:
\begin{equation}
\label{beta}
\beta(r) = 1 - \frac{\langle v_{\theta}^2 \rangle + \langle v_{\phi}^2
  \rangle}{2\langle v_r^2 \rangle} \,,
\end{equation}
where $\langle v_{\theta}^2 \rangle$, $\langle v_{\phi}^2
\rangle$ and $\langle v_r^2 \rangle$ are the mean square velocity
components in Galactic co-ordinates.

\subsubsection{Triaxial and anisotropic models}

We consider two anisotropic and/or triaxial halo models: the
logarithmic ellipsoidal model~\cite{logellip,evans} and the
Osipkov-Merritt model~\cite{osipkov,merritt}. We summarize these
models briefly below. For further details see either the original
papers~\cite{evans,osipkov,merritt} or Refs.~\cite{ullio:kamionkowski}
and ~\cite{evansomgreen}.

The logarithmic ellipsoidal model~\cite{logellip,evans} is the
simplest triaxial, scale free generalization of the isothermal sphere,
its shape and velocity anisotropy being independent of radius. This model
was studied in detail, in the context of WIMP direct detection, by
Evans, Carollo and de Zeeuw~\cite{evans} and has since been used
widely in calculations of the expected WIMP
signals~\cite{amevans,evansomgreen,amhalog2}. In this model, it is
assumed that the principle axes of the velocity distribution are
aligned with conical co-ordinates, and in any of the planes of the
halo conical co-ordinates coincide with local cylindrical polar
co-ordinates and the local WIMP distribution can be approximated by a
multi-variate Gaussian:
\begin{equation}
f_{0}(\vec{v}) = \frac{1}{(2\pi)^{\frac{3}{2}}
      \sigma_R\sigma_{\phi}\sigma_z}
       \exp\left(-\frac{v_R^2}{2\sigma_R^2}-\frac{v_{\phi}^2}
     {2\sigma_{\phi}^2}-\frac{v_z^2}{2\sigma_z^2}\right) \,,
\end{equation}
where $(\sigma_R,\sigma_{\phi},\sigma_{z})$ are the radial, azimuthal
and polar velocity dispersions respectively. These dispersions
are given in terms of
parameters $p$ and $q$ that are related to the halo axis ratios and
$\gamma$ which is related to the isotropy parameter
$\beta$~\cite{evans,evansomgreen}.

Osipkov-Merritt models~\cite{osipkov,merritt} provide self-consistent
radially anisotropic velocity distribution functions for halos with
spherically symmetric density profiles, with velocity anisotropy
$\beta(r) = r^2 /(r^2+r_{\rm a}^2)$ where $r_{\rm a}$ is the
anisotropy radius.  We follow Ref.~\cite{ullio:kamionkowski} and
assume that the MW has an NFW profile~\cite{NFW} with scale radius $20
\, $kpc~\footnote{The exact choice of profile has little influence on the
local velocity distribution~\cite{ullio:kamionkowski}; at the solar
radius the slope is close to $-2$ whatever profile is used.} and fix
$r_{\rm a} = 20$ and $12 \, {\rm kpc}$ as in Ref.~\cite{evansomgreen},
corresponding to $\beta(R_{0})= 0.14 $ and $0.31$, and calculate the
local velocity distribution using the fitting functions in
Ref.~\cite{widrow:fitting}.

The parameters of the fiducial halo models which we consider are
chosen to span the range of halo properties discussed above and are
summarized in Table~\ref{hm:param:def}. Some of the triaxial models
(numbers 6-9) are rather extreme (with $\sigma_{z} << \sigma_{r},
\sigma_{\phi}$), however they serve as a best/worse case scenario
allow us to assess whether the deviation of the recoil spectrum from
that produced by the standard halo is detectable/problematic.

\subsubsection{Tidal streams}
\label{stream}

The extent to which the local WIMP distribution is smooth is an open
question~\cite{stiff:widrow:frieman,moore:draco,helmi:white:springel,stiff:widrow2},
however it is certainly possible that a stream of high velocity
particles from a late accreted sub-halo could be passing through the
solar neighborhood~\cite{stiff:widrow:frieman,helmi:white:springel}
and would produce a potentially distinctive signal in direct detection
experiments~\cite{gondolo:sgr1}.  Indeed one of the tidal tails of the
Sagittarius dwarf galaxy (Sgr) passes close to the solar
neighborhood~\cite{mass:sgr,sdss:sgr}.  Helmi et al.~\cite{helmi}
previously discovered a group of halo stars in the solar
neighborhood moving coherently with mean velocity in Galactic
coordinates $\vec{v}_{\rm str} \approx (-65.0 ,\, 135.0 ,\, -250.0 )
\, { \rm km \, s}^{-1}$ and Freese et al.~\cite{gondolo:sgr2} have
argued that these stars are in fact part of the Sgr
tidal stream. This interpretation is somewhat controversial. Firstly (as
noted by Freese et al.)  there is a discrepancy between the measured
velocity of these stars and that expected for the Sgr stream.  
Secondly, there is a second lower density
stellar stream in the solar neighborhood moving in the opposite
direction~\cite{helmi} which is not expected for debris associated
with Sgr. Finally the comparison of detailed
simulations of the kinematics of the Sgr stream with measured
radial velocities of Sgr stream stars appears to indicate that
the stream does not pass through the solar
neighborhood~\cite{helmi2}.

None the less there is a debris stream passing through the solar
neighborhood regardless of whether or not it originated from the
Sgr dwarf galaxy.  We therefore follow
Ref.~\cite{gondolo:sgr1,gondolo:sgr2} and assume that there
is dark matter associated with the stream and model its velocity
distribution of WIMPs as a
maxwellian distribution with bulk velocity $\vec{v}_{\rm str}$, and
velocity dispersion $\sigma_{\rm str}$:
 \begin{equation} 
f_{0}(\vec{v})
= \frac{1}{(2\pi/3)^{3/2}\sigma_{\rm str}^3}\exp \left(
-\frac{3|\vec{v}-\vec{v}_{\rm str}|^2}{2\sigma_{\rm str}^2} \right) \,.
\end{equation} 
We take $\sigma_{\rm str} = 30 \, {\rm km \, s}^{-1}$,
and $\rho_{0} = 0.07 \, {\rm GeV \, cm}^{-3} (\approx 0.25 \rho_{0})$, 
(towards the lower and upper ends
of the range of values suggested in Ref.~ \cite{gondolo:sgr1,gondolo:sgr2} 
respectively
and therefore providing a best case scenario for detectability), and for
simplicity use the standard maxwellian velocity distribution [eq.~(\ref{fmax})] for the smooth background component.

\begin{table}[t!]
\begin{tabular}{|c|c|c|c|c|c|c|c|}
\hline
No. & Type & p & q & $\beta$ & axis &  $\rho_{\rm str} / \rho_{0}$
 & $\sigma_{\rm str}$ (km \, s$^{-1}$)\\
\hline
1 & SHM & - & - & - & - & - & -\\
2 & LGE & 0.9 & 0.8 & 0.1 & I & - & -\\
3 & LGE & 0.9 & 0.8 & 0.4 & I & - & -\\
4 & LGE & 0.9 & 0.8 & 0.1 & L & - & -\\
5 & LGE  & 0.9 & 0.8 & 0.4 & L & - & -\\
6 & LGE & 0.72 & 0.7 & 0.1 & I & -& -\\
7 & LGE & 0.72 & 0.7 & 0.4 & I & - & -\\
8 & LGE & 0.72 & 0.7 & 0.1 & L & - & -\\
9 & LGE  & 0.72 & 0.7 & 0.4 & L & - & -\\
10 & OM & - & - & 0.31 & - & - & -\\
11 & OM & - & - & 0.14 & - & - & -\\
12 & SHM.+Str  & - & - & - & - & 0.25 &  30 \\
\hline
\end{tabular}
\caption{Summary of the parameters of the fiducial halo models. `SHM' 
denotes the standard halo model,
`LGE' the logarithmic ellipsoidal model with the Sun located on either 
the long (L) or intermediate (I) axis, `OM' 
the Osipkov-Merritt model and `SHM+ Str' the standard halo 
model plus tidal stream contributing
$ \rho_{\rm str}/\rho_{0}$ of the local WIMP density, 
with bulk velocity and velocity dispersion as described in the text.
In each case
we take $v_{{\rm c}} = 220 \, {\rm km \,s}^{-1}$ and 
$\beta$ is the anisotropy parameter, defined as in eq.~(\ref{beta}).}
\label{hm:param:def}
\end{table}

\subsection{Modeling the detector response}
\label{detector}

TPC based detectors are designed to measure the directions of nuclear
recoils by drifting the ionisation produced by recoils in the gas
volume to a suitable charge readout plane. The difference in ionisation track
lengths between electrons, alpha particles and nuclear recoils at low gas
pressure allows efficient rejection of these
backgrounds~\cite{drift}. To reduce
charge diffusion along the track (which would
restrict the spatial resolution and hence direction measurement and
background discrimination) to thermal levels the detector can be
filled with an electronegative gas to give negative ion drift. The
DRIFT collaboration have adopted CS$_2$ as a suitable electronegative
gas; for further details see Ref.~\cite{drift,sean:drift}. 

The recoiling nucleus undergoes a large number of scatterings (on
average 30 (70) for a primary recoil energy of 20 (100) {\rm keV}, with
a gaussian distribution).
These multiple scatterings  together with the small,
but finite, diffusion of drifted ionisation limits the accuracy with
which the direction of the primary recoil can be reconstructed.  In
order to assess the likely accuracy of the track reconstruction we
consider a TPC detector filled with 0.05 bar CS$_2$ with a 200$\, \mu
$m pitch micropixel readout plane, a 10cm drift length over which a
uniform electric field of $1\, {\rm kV \, cm}^{-1}$ is applied. This
gas pressure, mixture and electric field are chosen to match the
design of the DRIFT-I detector~\cite{sean:drift}. The micropixel
readout is based on those described in Ref.~\cite{bellazini}, and the
drift length is such that the rms diffusion over the full drift length
is approximately equal to the pixel pitch~\cite{ohnuki}.  We use the
SRIM2003 package~\cite{srim}~\footnote{This package was not designed
to model recoils in gaseous targets, however it accurately reproduces
the recoil ranges in S found experimentally.} to generate sulfur
ionisation recoil tracks~\footnote{Carbon recoils are ignored as they
account for less than 1 $\%$ of the recoil rate due to the $A^2$
dependence of the recoil rate and the low Carbon mass fraction.}  We
simulate the drift and diffusion of the ionisation to the readout
plane under the electric field and the subsequent generation of charge
avalanches.    The quenching factor
(fraction of recoil energy going into ionization) is roughly 0.4 for
a primary recoil energy of 20 keV, in agreement with
experimental data~\cite{quench}, however the direction reconstruction
depends on the electron distribution and not their number.

\begin{figure}[t!]  \includegraphics[width=8.5cm]{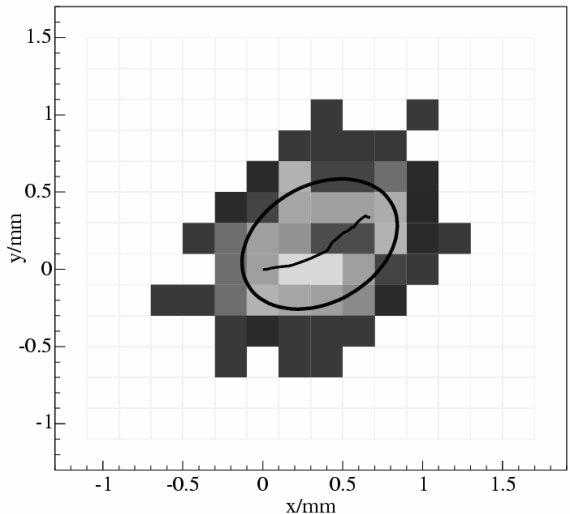}
 \caption{Simulated pixel read-out in the xy plane from a 40keV recoil.
The strength of the signal in each pixel
is obtained by integrating the voltage pulse for that pixel. The actual
recoil track is shown by the solid line. The ellipse  results
from the moment analysis and its long axis gives the 
reconstructed direction.}
\label{recon}
\end{figure}

The charge distributions are then projected into the xy, yz and xz
planes (as illustrated in Fig.~\ref{recon}). The resulting pixel hit
pattern is approximately elliptical with long axis close to the
primary recoil direction, which allows us to reconstruct the nuclear
recoil directions via a moment analysis. The distribution of the
difference between the primary recoil direction and the reconstructed
track direction peaks at $\sim 15^{\circ}$, decreasing weakly with
increasing energy, and has a long large angle tail (the mean and root
mean square deviation are $\sim 25^{\circ}$ and $\sim 15^{\circ}$,
again weakly decreasing with increasing energy).  We include this
stochastic uncertainty in our Monte Carlo simulations.

The primary recoil energy threshold is 20 keV as below
this energy the short track length (~3-4 pixels)
and multiple scatterings make it
impossible to reconstruct the track direction. With
SRIM2003 generated recoils, we find near uniform distributions of
ionisation along the tracks and therefore cannot determine the
absolute signs of the reconstructed recoil vectors (i.e $+\vec{x}$ or
$-\vec{x}$). However, we note that experimental measurements are
really required to determine whether this absolute sign can be
measured or not.  We consider both possibilities below.

We assume that the time-dependent conversion of recoil directions
measured in the laboratory frame to the galactic frame introduces no
further errors in the measured recoil directions due to, for instance,
inaccuracy in the measured time of the event.

\vspace{0.5cm}

We calculate the directional recoil rate above the $20 \, {\rm keV}$
detector energy threshold via Monte Carlo simulation so as to allow
the inclusion of the detector resolution effects discussed above. We
work in Galactic $(R, \phi, z)$ co-ordinates, where $R$ is directed
towards the Galactic center, $\phi$ is in the direction of motion of
the Local Standard of Rest (LSR) and $z$ points towards the north
Galactic pole. These co-ordinates are related to Galactic longitude
$l$ and latitude $b$ by $(R,\, \phi, \, z)=(\cos{l} \cos{b} , \sin{l} 
\cos{b}, \sin{b})$.

Firstly the WIMP velocity distribution has to be transformed to the
rest frame of the detector $[f_{0}(\vec{v}_{\chi}) \rightarrow
f_{\oplus}(\vec{u}_{\chi})]$ via the Galilean transformation:
$\vec{v}_{\chi} =\vec{u}_{\chi}+\vec{v}_{\oplus}(t)$. The velocity of
the Earth, $\vec{v}_{\oplus}(t)$, is the sum of the velocity of the
LSR, $\vec{\Theta}_{LSR}$, the peculiar velocity of the Sun,
$\vec{v}_{\odot p}$, and the Earth's orbital velocity about the Sun,
$\vec{v}_{orb}(t)$. We take $\vec{\Theta}_{LSR}= 220 \, {\rm km \,
s}^{-1}$ (see e.g.~\cite{binney:tremaine}), $\vec{v}_{\odot p}=(10.0,
7.3, 5.2) \, {\rm km \, s}^{-1}$~\cite{dehnen:binney:hipparcos}, and
use the results of the calculation of the Earth's orbital velocity
detailed in Appendix A.  The resulting WIMP velocity distribution in
the detector rest frame is used to generate random incident WIMP
velocities from the WIMP flux $ u_{\chi} f_{\oplus}(\vec{u}_{\chi})$, which
then undergo isotropic scattering in the center of mass frame. We
weight each recoil by a factor equal to the detector form factor
(which is taken to have the Helm form appropriate for spin independent
WIMP-sulfur scattering~\cite{ls}) in order to take into account the
variation of the scattering cross section with the momentum transfer
to the nucleus.  Finally we include the effects of multiple scattering
on the reconstructed recoil direction.

\begin{figure}[t!]  \includegraphics[width=8.5cm]{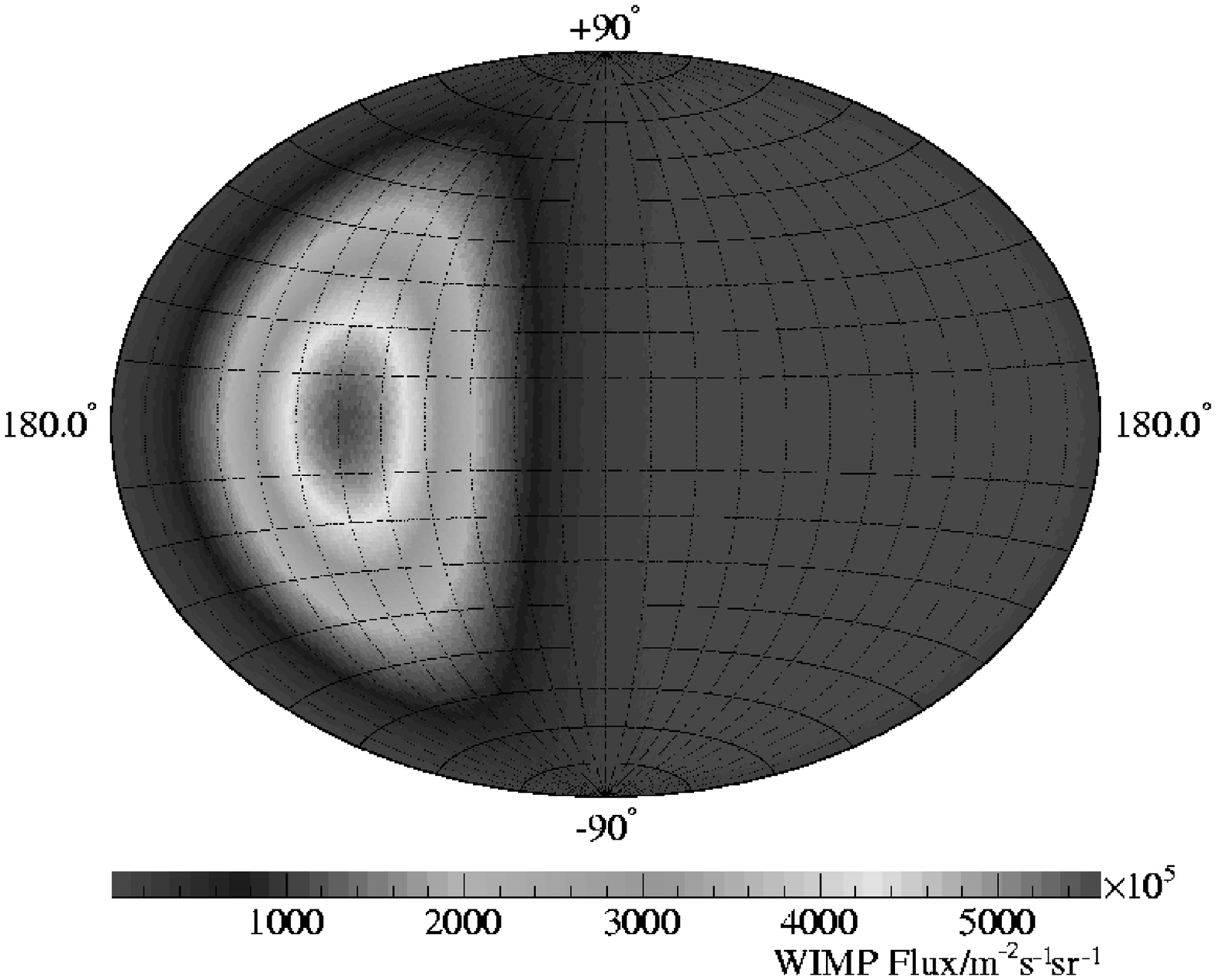}
\includegraphics[width=8.5cm]{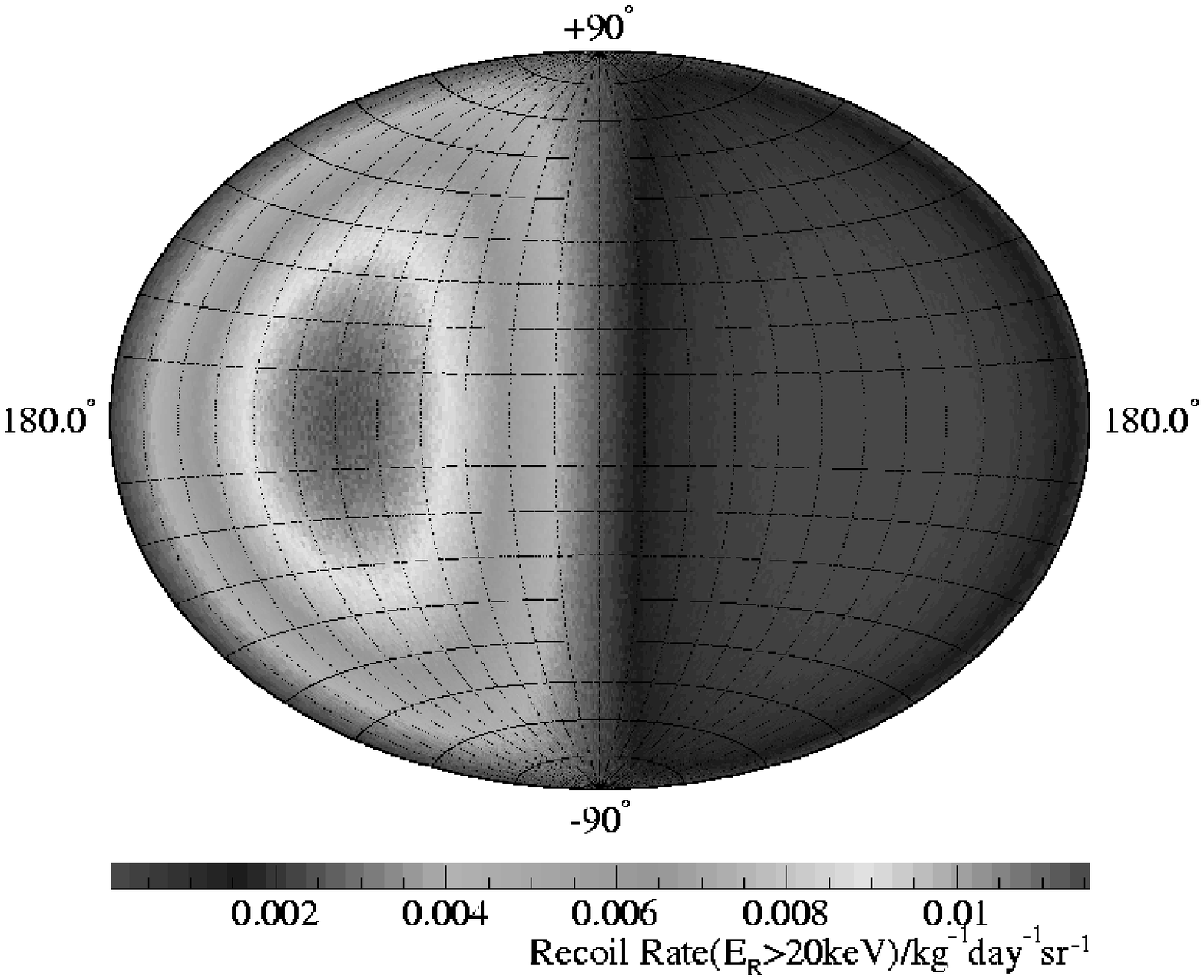} \caption{Time
averaged WIMP flux (top) and S Recoil rate above
$20 \, {\rm keV}$ (bottom) in Galactic
$(l,b)$ co-ordinates for halo
 model 1 and $\rho_0=0.3 \, {\rm GeV \, cm}^{-3}$, $m_{\chi}=100$ GeV 
and  $\sigma_{\chi p}=10^{-6}$ pb.}
\label{skyplot:1}
\end{figure}

\begin{figure}[t!]
\includegraphics[width=8.5cm]{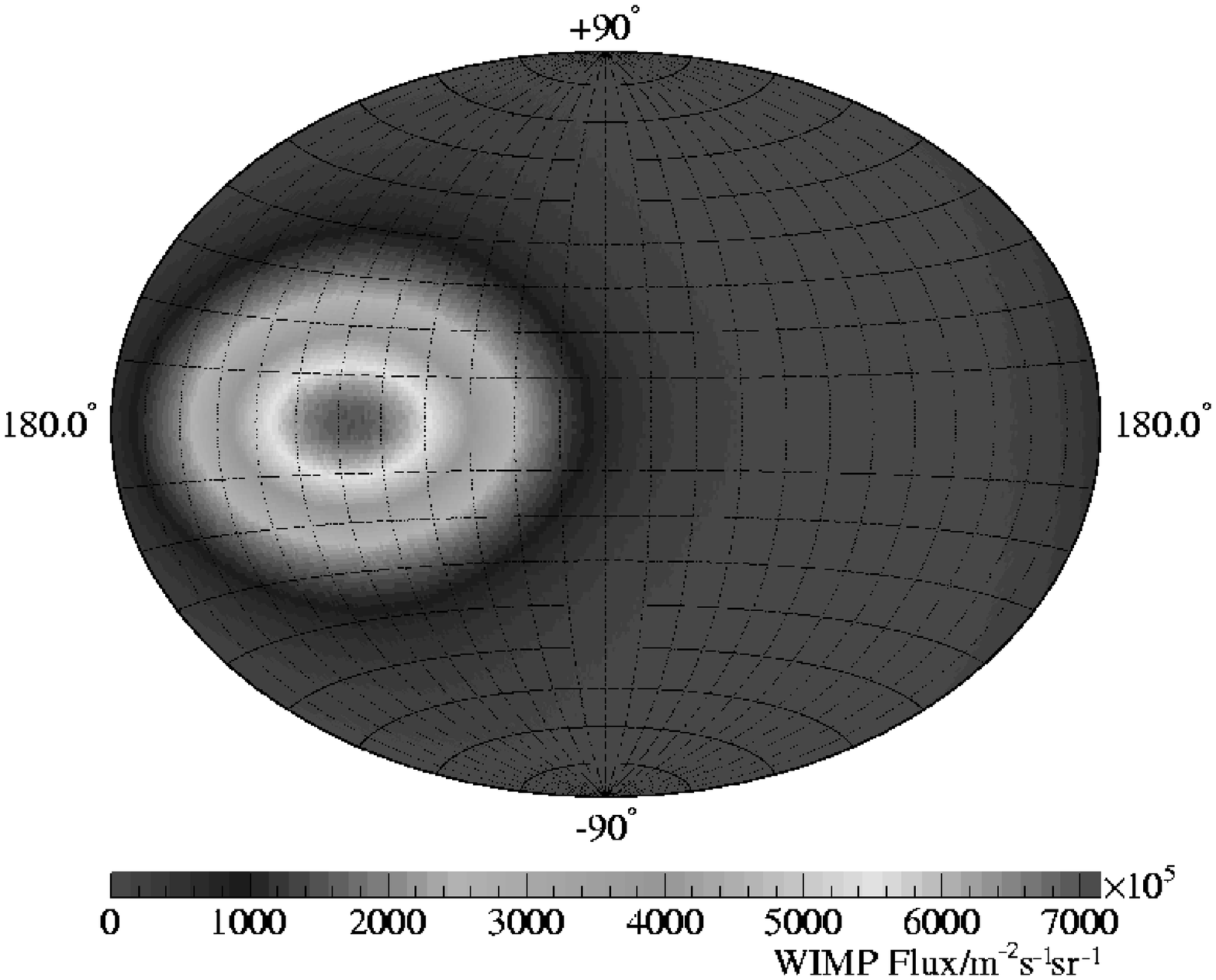}
\includegraphics[width=8.5cm]{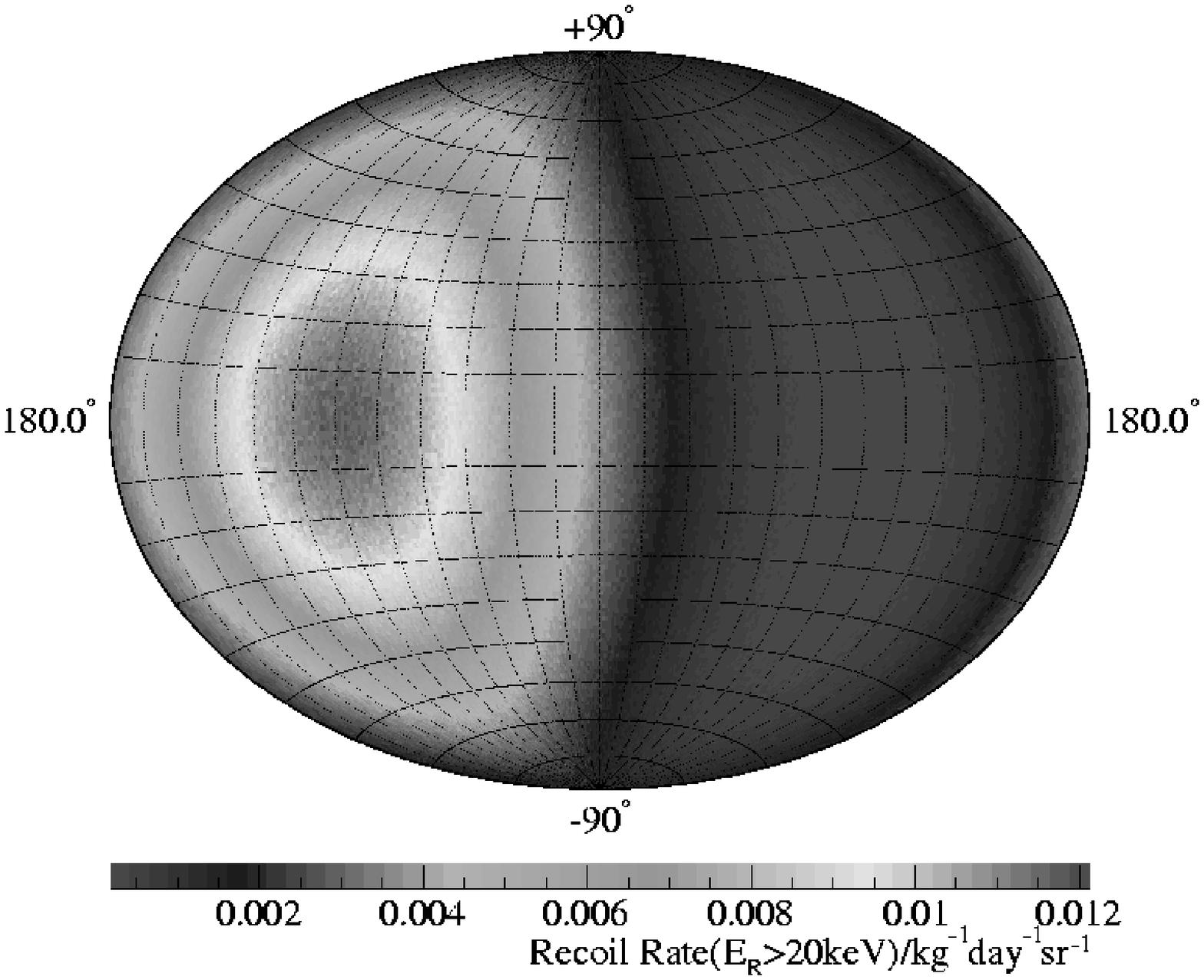}
\caption{As Fig.~\ref{skyplot:1} for model 3 (logarithmic ellipsoidal 
model with $p=0.9, q=0.8$ and
$\beta=0.4$ ).}
\label{skyplot:3}
\end{figure}

\begin{figure}[t!]
\includegraphics[width=8.5cm]{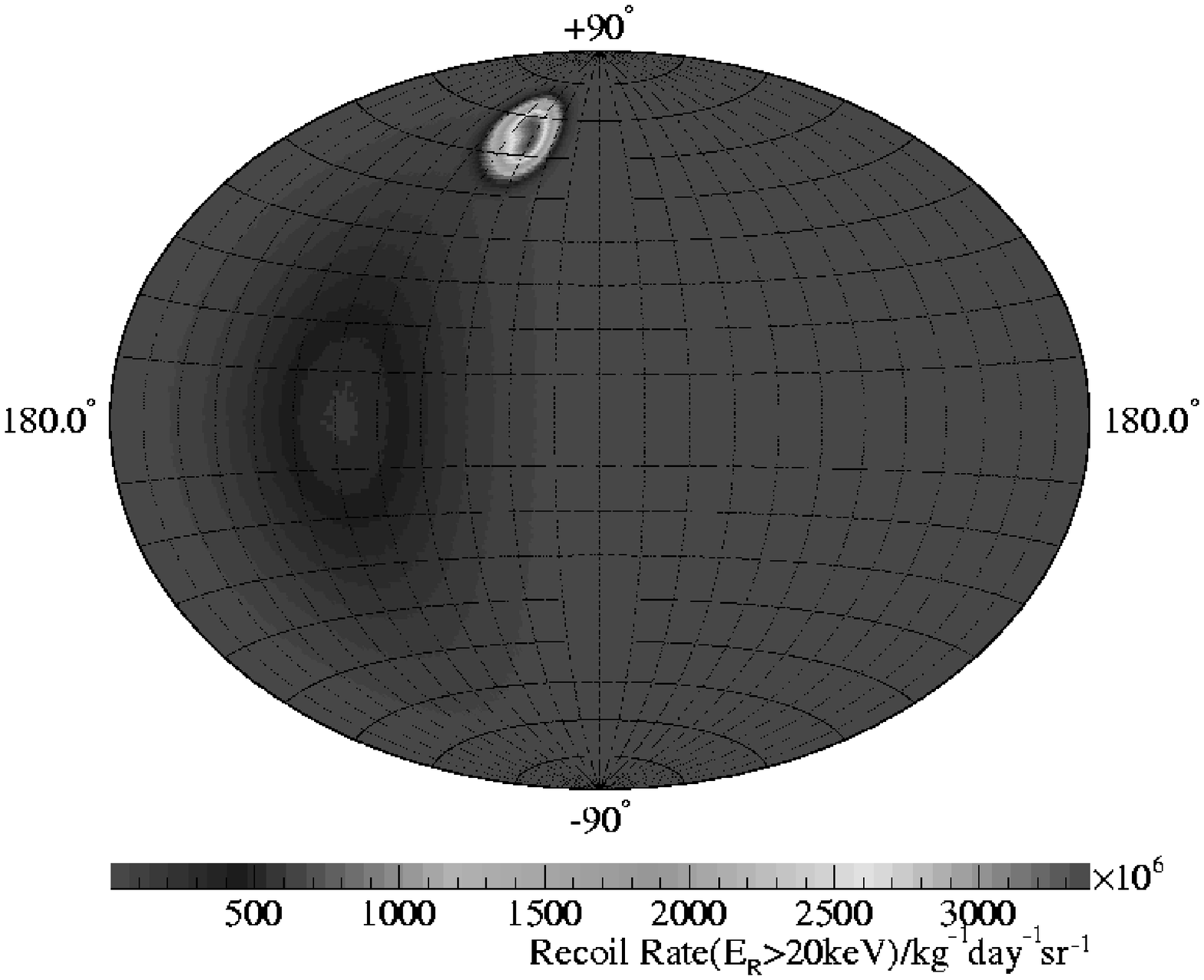}
\includegraphics[width=8.5cm]{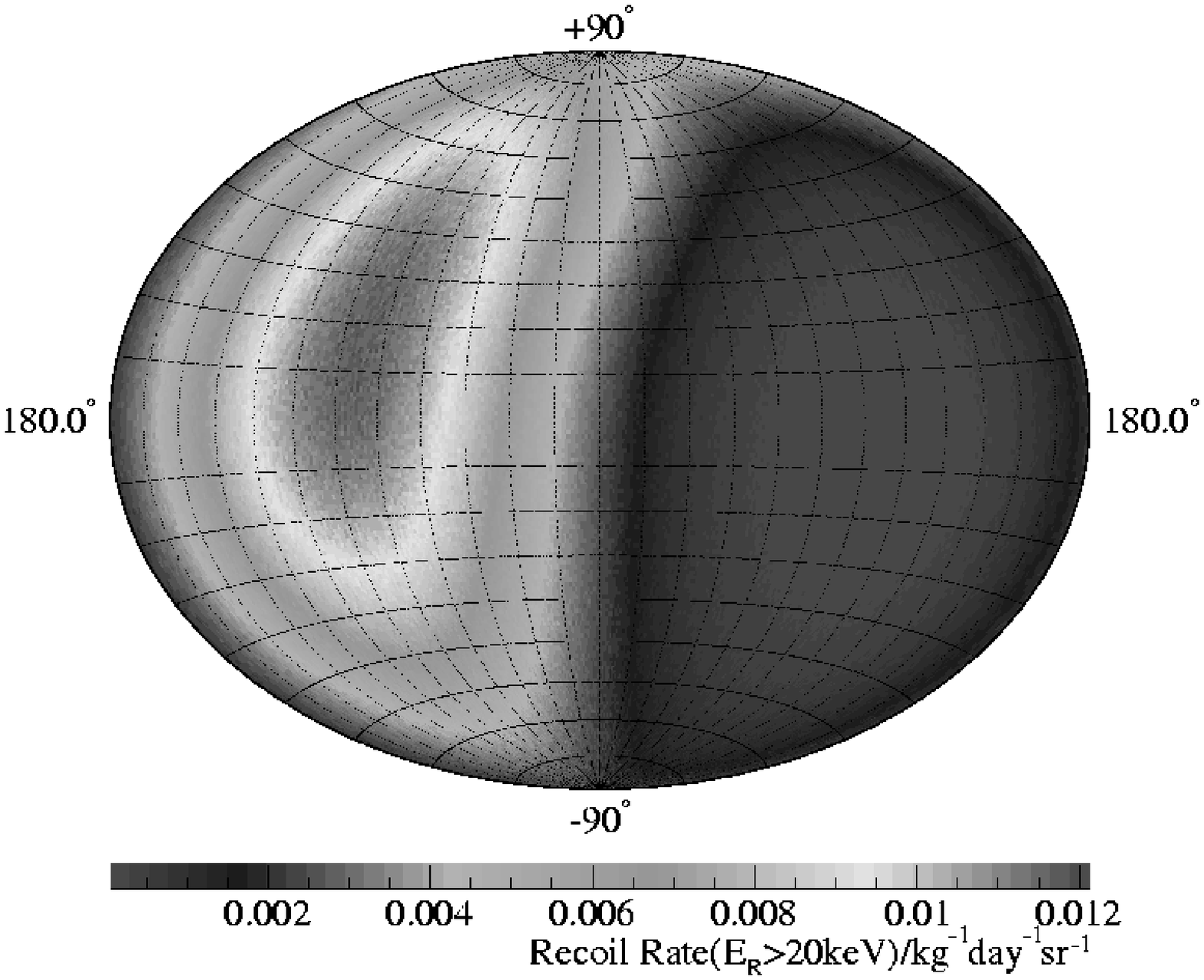}
\caption{As Fig.~\ref{skyplot:1} for model 12 (standard halo model 
with a 25$\%$ contribution
to the local WIMP density from a tidal stream with properties given in 
Sec.~\ref{stream}).}
\label{skyplot:12}
\end{figure}

The resultant, time averaged, WIMP and recoil distributions are shown
as Hammer-Aitoff projections of the flux/rate in Galactic coordinates
in Figs.~\ref{skyplot:1}-\ref{skyplot:12} for models 1 (standard halo
model), 3 (logarithmic ellipsoidal model with $p=0.9, q=0.8$ and
$\beta=0.4$ ) and 12 (standard halo model with a 25$\%$ contribution
to the local WIMP density from a tidal stream with properties as described
in Sec.~\ref{stream}).
For illustrative purposes we take $\rho_0=0.3 \, {\rm GeV\, cm}^{-3}$,
$m_{\chi} = 100 \, {\rm GeV}$ and for the recoil flux take the
WIMP-proton elastic scattering cross-section to be $\sigma_{\chi
p}=10^{-6} \,$pb. Note the different scale for model 12.
We see that the WIMP flux produced by the triaxial halo
model is significantly flattened with respect to the Galactic plane,
however due to the finite detector resolution the difference between
the recoil rates predicted by models 1 and 3 is far smaller than the
difference in the WIMP fluxes. For model 12, with the 
tidal stream, the peak direction of the WIMP flux and recoil rate both
differ significantly from the direction of motion of the Sun,
suggesting that WIMPs associated with the stream of halo stars
in the solar neighborhood could be detectable
in a directional detector if the WIMP density is sufficiently large.

\section{Applying Statistical tests}
\label{stat} 

A WIMP search strategy with a directional detector can essentially be
divided into three regimes. Firstly, there is the simple search phase
which aims to detect an anomalous recoil signal above that expected
from backgrounds. Secondly,
following the discovery of an anomalous recoil signal, there would
be the confirmation stage. At this point the experiment would
collect more data and aim to confirm the recoil signal as Galactic
in origin by searching for the expected anisotropy in the recoil
directions. In statistical terms the question posed at this point is
`Is the distribution of observed recoil directions isotropic?', more
specifically `How many events are required to reject the hypothesis
of isotropy (at a given confidence level)?'.  Thirdly, assuming
anisotropy is detected, the experiment would then collect more data
and try and determine the form of the recoil anisotropy and hence
the local WIMP velocity distribution. The halo models considered in
Sec.~\ref{halo} lead us to consider two simple hypotheses to test:
 1) Does the distribution of recoil directions show evidence of
 flattening (i.e. a non-spherical halo)? and 2) Does the distribution
 of recoil directions show evidence for a signal from a tidal stream?

As discussed in Sec.~\ref{intro}, previous work aimed at determining
the number of events required to detect the anisotropy in the recoil
directions has indicated that $5-30$ events are required to give a
90-95\% rejection of isotropy~\cite{copi:krauss,lehner:dir}. However,
these analyses suffer from several non-optimal features.  Neither
analysis took into account the angular resolution of the recoil
reconstruction or the possibility that the absolute signs of the
recoil vectors (i.e. their `sense' $+\vec{x}$ or $-\vec{x}$) may not
be measurable.  In Ref.~\cite{lehner:dir}, a Kolmogorov-Smirnov
goodness of fit test was applied to the binned distribution of recoil
$\cos\gamma$ values, where $\gamma$ is the angle between the recoil
direction and the direction of motion of the Sun. Given the small
number of events expected at the confirmation stage of an experiment,
binning data can lead to a significant loss of information and should
therefore be avoided. Ref.~\cite{copi:krauss} used an unbinned
likelihood analysis to determine the number of events required for a
95\% confidence detection of isotropy in 95\% of experiments, for a
range of halo models, and also the number of events required to
distinguish a triaxial halo from the standard maxwellian halo
model. The draw-back of likelihood analysis is that it requires the
likelihood function for the recoil distribution generated by a given
halo model. It can therefore only give the relative likelihood of
specific models (and parameter values), and there is no guarantee that
any of the halo models considered is a good approximation to the local
WIMP velocity distribution (especially since non-spherical and/or
anisotropic halo models involve assumptions, which may or not be
valid). As such this analysis could not be applied to real data.

Recoil directions constitute vectors, or, if the senses are unknown,
undirected lines or axes, and so can equivalently be represented as
points on a sphere. This allows us to use statistical inference
methods developed for the analysis of spherically distributed data
(for a review of this extensive field see the standard texts such
as~\cite{mardia:jupp,fisher:lewis:embleton}).  We investigate a
variety of non-parametric statistics designed to test the isotropy
(Secs.~\ref{iso}), rotational symmetry (Sec.~\ref{rotsymtest}) and
median direction (Sec.~\ref{dirtest}).  When quoting results for the
numbers of events required for detection, we focus on the benchmark
halo models discussed in Sec.~\ref{halo}, however these tests do not
make any assumptions about the form of the recoil spectrum and are
hence equally applicable to any local WIMP velocity
distribution. Recoil energies are not incorporated in these
statistics, but the energy provides no extra information about the
degree of anisotropy and/or rotational symmetry in the data. In
addition, any analysis including the recoil energy would require (halo
model dependent) assumptions about the form of the recoil energy
spectrum as a function of direction. 

Throughout we assume zero background. This is a reasonable expectation
for experiments such as DRIFT-II made from low activity materials with
efficient gamma rejection and shielding~\cite{drift2:design}.
However, non-zero backgrounds can be incorporated into the statistical
tests. For real data, one would simply calculate the value of the
appropriate test statistic using the observed sample of recoil
vectors/axes and use its null distribution (Appendix B1) to determine
the probability of the sample being isotropic. No assumption about the
level of background contribution to the sample is neccessary, but
there is an implicit assumption that background event directions are
isotropically distributed.

In the case of our simulations, it would be neccesary to fix the
background rate and exposure so that the expected number of background
events $N_b$ is known. A variable number of WIMP events $N_{\rm w}$
can then be added to the $N_{\rm b}$ background events, and the
formalism of Appendix C used to determine the value of $N_{\rm w}$
required to reject isotropy of the $N_{\rm w}+N_{\rm b}$ recoil sample
at a given confidence level. Since the background rate and exposure
are dependent on the experiment, investigating the effect of varying
background rates is beyond the scope of this paper. Therefore, and
given that the next generation of directional detectors are expected
to have essentially zero background~\cite{drift2:design}, we assume
zero background to provide benchmark figures.  Nevertheless, the
formalism presented in this paper could be used to predict the
directional sensitivity of a detector with non-zero background once
the background rate and exposure have been estimated.

\subsection{Tests of isotropy}
\label{iso}

The first fundamental question posed by a directional detector
observing an anomalous recoil signal is `Are the observed recoil
directions consistent with an isotropic distribution?'. The most
general tests of isotropy are those which do not depend on the
coordinate system in which the sample vectors/axes are measured. This
independence from the coordinate system means no assumption about the
form of the anisotropy in the signal is required.
While this is usually an advantage, 
coordinate independent tests cannot make use of any information that is
available about the expected form of the anisotropy. If the local WIMP
distribution is smooth then the year averaged recoil flux will be peaked in
the direction of the Sun's motion at
$(l_{\odot},b_{\odot})=(87.5^{\circ},1.3^{\circ})$.

For each of the statistics discussed in Appendix~\ref{iso:A} we
determine the minimum number of events required to reject isotropy of
recoil directions at $90 \, (95) \%$ confidence in $90 \, (95)\%$ of
experiments (i.e. for rejection and acceptance probabilities of $
R_{\rm c} =A_{\rm c}=0.9$ and $0.95$), $N_{\rm iso}$, as described in
Appendix~\ref{hyptest}, for each statistic and halo model considered,
including the detector response in the calculation of the recoil
distributions.  For the standard maxwellian halo model we also carry
out the calculation neglecting the reconstruction uncertainty and for
a hypothetic perfect detail which also has zero energy threshold , in
order to assess the effect of the reconstruction uncertainties and the
finite energy threshold on the number of events required. The results
are tabulated in Table~\ref{nevents} for each of the fiducial halo
models for each statistic.

\begin{table}[htbp]
\begin{tabular}{|c|c|c|c|c||c|c|c|}
\hline
Halo & \multicolumn{7}{c|}{ $N_{\rm iso}$ for
 $( R_{\rm c}, A_{\rm c})=(0.90,0.90)$} \\
\cline{2-8}
Model & \multicolumn{4}{c||}{Vectorial Statistics} & \multicolumn{3}{c|}{Axial
 Statistics} \\
\cline{2-8}
& ${\cal W}^{\star}$ & ${\cal A}$ & ${\cal F}$ & $\langle \cos \theta \rangle$ &
${\cal B}^{\star}$ & ${\cal G}$ & $\langle |\cos \theta| \rangle$ \\
\hline
1 & 12 & 12 & 13 & 7 & 165 & 167 & 81 \\
1 (no) & 10 & 10 & 10 & 5 & 83 & 84 & 40 \\
1 (per) & 18 & 18 & 18 & 10 &531 &538 & 258 \\
2 & 12 & 12 & 12 & 7 & 114 & 114 & 57 \\
3 & 14 & 14 & 15 & 8 & 157 & 159 & 93 \\
4 & 12 & 12 & 13 & 7 & 149 & 151 & 74 \\
5 & 14 & 14 & 15 & 8 & 157 & 159 & 93 \\
6 & 11 & 11 & 11 & 6 & 67 & 67 & 36 \\
7 & 14 & 14 & 14 & 8 & 88 & 90 & 57 \\
8 & 13 & 13 & 14 & 7 & 175 & 178 & 86 \\
9 & 15 & 15 & 16 & 9 & 264 & 267 & 146 \\
10 & 15 & 15 & 16 & 8 & 280 & 284 & 149 \\
11 & 12 & 12 & 12 & 7 & 125 & 126 & 62 \\
12 & 14 & 14 & 14 & 8 & 221 & 222 & 159 \\
\hline \hline
 & \multicolumn{7}{c|}{$N_{\rm iso}$ for
 $(R_{\rm c}, A_{\rm c})=(0.95,0.95)$} \\
\hline
1 & 18 & 18 & 19 & 11 & 235 & 235 & 131 \\
1 (no) & 13 & 13& 14 & 9 & 120 &  120 & 65 \\
1 (per) & 25 & 25  & 26 & 15  & 767    &  769   & 426 \\
2 & 17 & 17 & 18 & 10 & 161 & 163 & 93 \\
3 & 20 & 20 & 21 & 12 & 225 & 224 & 152 \\
4 & 18 & 18 & 19 & 11 & 215 & 214 & 120 \\
5 & 20 & 20 & 21 & 12 & 222 & 22 & 153 \\
6 & 16 & 16 & 16 & 10 & 96 & 97 & 59 \\
7 & 19 & 20 & 20 & 12 & 125 & 125 & 93 \\
8 & 18 & 18 & 19 & 11 & 252 & 252 & 142 \\
9 & 21 & 21 & 22 & 13 & 378 & 383 & 241 \\
10 & 21 & 21 & 21 & 12 & 402 & 405 & 247 \\
11 & 17 & 17 & 17 & 10 & 180 & 182 & 102 \\
12 & 19 & 19 & 20 & 12 & 316 & 320 & 264 \\
\hline
\end{tabular}
\caption{Number of recoil events
required to reject isotropy of recoil directions, 
$N_{\rm iso}$, at $90  \, (95)\%$
confidence in $90 \, (95)\%$ of experiments 
for each test statistic and
halo model considered, including the uncertainty in the nuclear 
recoil reconstruction
and for the standard halo model 
ignoring the uncertainty in the nuclear recoil reconstruction
(denoted by `no' in the table) and
for a hypothetical perfect detector with zero threshold and no reconstruction
uncertainities (denoted by `per').
The test statistics have been divided into those suitable
for vectorial data  and those applicable to
axial data (i.e. when the senses of the recoil directions are not known).}
\label{nevents}
\end{table}

Overall, the two coordinate dependent tests of isotropy, $\langle \cos
\theta \rangle$ and $\langle |\cos \theta| \rangle$, have the lowest
$N_{{\rm iso}}$ values and are thus the most powerful tests for
rejecting isotropy. All the coordinate independent tests typically
require 1.5-2 times more events to reject isotropy for a given halo
model (for both vectorial and axial data). This is perhaps not
surprising for the models 1-11 however, the coordinate dependent tests
are also the most powerful for rejecting isotropy in model 12 which
includes a $25\%$ contribution to the local WIMP density from a tidal
stream.

The vectorial coordinate independent tests (${\cal
W}^{\star}$, ${\cal A}$ and ${\cal F}$) all have very similar values
of $N_{{\rm iso}}$ and are hence equally powerful for rejecting
isotropy of recoils.
The same is true of the axial coordinate independent tests (${\cal
B}^{\star}$ and ${\cal G}$), however $N_{{\rm iso}}$ is of order ten
times larger for these tests than for the vectorial tests, and there
is much greater variation between the fiducial halo models.  Both of
these effects are due to the form of the recoil direction
distribution and the way in which lack of knowledge of the recoil
sense changes this form. While most of the recoil arrival directions
lie in the forward hemisphere, $0<l<180^{\circ}$, a
non-negligible (halo model dependent) number lie in the backward
hemisphere. For the vectorial tests, these backward events are not a
problem as the tests can distinguish between the forward and backward
hemispheres. When only the recoil axis is known, events in the
backward hemisphere are effectively `added' to the forward
hemisphere, reducing the degree of anisotropy.  
Halo model 6 has the smallest rate in the backward
hemisphere and thus the lowest $N_{{\rm iso}}$ for both vectorial and
axial tests, while model 10 has the largest backward rate and
therefore has higher $N_{{\rm iso}}$ values with axial tests.  In
general the broader the local WIMP velocity distribution, the greater
the number of events in the backward hemisphere, and hence a larger
number of events are required to reject isotropy if the sense of the
recoil directions are not known.

Ignoring the detector response only marginally reduces the number of
events required for the vectorial tests, however the number of events
required for the axial tests are decreased by a factor of roughly two.
For the hypothetical detector with perfect reconstruction and zero
energy threshold the number of events required increases, relative to
the 20 keV perfect reconstruction case, by a factor of $\sim 2 (\sim
6)$ for the vectorial (axial) statistics. The number of events
required increases since, as is well known~\cite{dirndep}, the
anisotropy of the recoil directions decreases with decreasing recoil
energy. The total recoil rate with a 0 keV threshold is roughly twice
that with a 20 keV threshold, therefore for the vectorial statistics
the exposure required to detect a WIMP signal is essentially
unchanged, while for the axial statistics (i.e. if it is not posible
to measure the senses of the recoils) the lower threshold would
actually increases the exposure required.

We now translate the number of events required to reject isotropy for
the realistic detector into the equivalent detector exposure, $E$,
required to observe this number of events. If the senses of the recoil
directions are observed, isotropy could be rejected at $95\%$ confidence
in $95\%$ of experiments for $\sigma_0 \sim 3\times10^{-9} \,{\rm pb}$
and $\rho_0\sim0.3 \, {\rm GeV \, cm}^{-3}$ with an exposure of $E
\sim 10^5 \, {\rm kg \, day}$ (i.e. a 100 kg detector operating for a
period of 2-3 years). For a detector only capable of measuring the
recoil axes a $10^5 \, {\rm kg \, day}$ exposure would be able to
reject isotropy at the same confidence level and acceptance down to
$\sigma_0\sim3\times10^{-8} \, {\rm pb}$.

\subsection{Tests for rotational symmetry}
\label{rotsymtest}

Once an observed anomalous recoil signal is found to be incompatible
(at some confidence level) with isotropy the next question to pose is
`Is the observed distribution consistent with a spherical halo?'.  A
generic feature of triaxial halo models is a flattening of the recoil
distribution towards the galactic plane, or more generally, flattened
along one principal axis of the local velocity distribution.  This
type of distribution can be probed using tests for rotational symmetry
about a specified direction or axis. In the case of smooth halo models
this direction/axis is the direction of motion of the Earth through
the MW halo, which over the year, averages to the direction of motion
of the Sun, $(l_{\odot},b_{{\odot}})$~\footnote{For experiments with
an exposure that is non-uniform in time, the mean direction could be
calculated by averaging the Earth's velocity vector over the,
time-dependent, exposure.}.

We focus here on the two fiducial halo models for which the recoil
distribution deviates most from that produced by the standard halo
model: the logarithmic ellipsoidal models 5 ($p=0.9$, $q=0.8$,
$\beta=0.4$, long axis) and 7 ($p=0.72$, $q=0.7$, $\beta=0.4$,
intermediate axis) ~\footnote{More spherical models that produce less
flattened recoil distributions would require more events to
reject rotational symmetry.} and find the number of events required to
reject rotational symmetry, $N_{{\rm rot}}$, using the Kuiper statistic
${\cal V}^{\star}$ as defined in Appendix~\ref{rotsymtest:A}. We find
that halo model 5 has $ A<0.4$ over a large range of $N$ values, and
of order 5000-8000 events would be required to reject rotational
symmetry at $90+ \%$ confidence with similar acceptance.  In contrast
for halo model 7 (which is more extreme) $N_{{\rm rot}} = 1170 \,
(1710)$ for $R_{\rm c}=A_{\rm c} = 0.9 \, (0.95)$.  So while
rotational symmetry of the recoil distribution can be rejected at high
confidence and acceptance for this, rather extreme, halo model, it
requires 10-50 times the number of events required to reject isotropy
(depending on whether vectors or axes are measured in the latter
case).

Converting these numbers to the equivalent exposures in ${\rm kg
\,day}$ gives $E(R_{\rm c}=A_{\rm c}=0.9 \, (0.95)) = 7.8 \, (11)
\times10^{-3} \sigma_0^{-1}\rho_0^{-1}$.  Hence for the $10^5 \, {\rm
kg \, day}$ exposure considered in the previous subsection, rotational
symmetry could be rejected at $90\%$ confidence down to $\sigma_0\sim
3\times10^{-7} \, {\rm pb}$ for $\rho_0\sim0.3 \, {\rm GeV \,
cm}^{-3}$, independent of whether the senses of the recoils can be
measured. The sensitivity of a TPC-based directional detector for
rejecting a spherical halo if the real MW halo is significantly
triaxial using this statistic, is therefore several orders of
magnitude lower than for rejecting isotropy of recoil directions.
Devising other tests that may be a more powerful probe of the
flattening of the recoil distribution produced by non-spherical halo
models is therefore an important task.

\subsection{Tests for mean direction}
\label{dirtest}

\begin{figure}[t!]
\includegraphics[width=8.5cm]{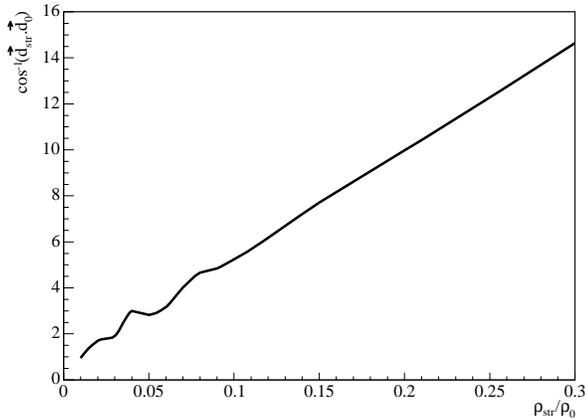}
\caption{Variation
  of the angle (in degrees) between the mean recoil directions 
of the standard
  halo and the standard halo plus a tidal stream with 
$\vec{v}_{\rm str} = (-65.0 ,\, 135.0 ,\, -250.0 )
\, { \rm km \, s}^{-1}$
  function of the fraction of the local WIMP density contributed by 
the stream, $\rho_{\rm str}/ \rho_{0}$.}
\label{sgr:angle}
\end{figure}

Tidal streams will typically have a velocity dispersion which is small
compared to their velocity relative to the solar system and therefore
the recoil distribution due to WIMPs from a stream will be peaked in
the hemisphere whose pole points in the direction of the stream
velocity.  The net (stream plus smooth background WIMP distribution)
peak direction depends on the fraction of the local density that the
tidal stream contributes.  Fig.~\ref{sgr:angle} shows the angle
between the direction of the Sun's motion
$(l_{\odot},b_{\odot})=(87.5^{\circ},1.3^{\circ})$ and the direction
of the peak in the recoil distribution produced by a maxwellian halo
plus a stream with velocity $\vec{v}_{\rm str} = (-65.0 ,\,
135.0 ,\, -250.0 ) \, { \rm km \, s}^{-1}$ as a function of the local
WIMP density contributed by the stream. For the $\rho_{\rm str}/\rho_0
\sim 1-30\%$ range of stream densities suggested
by~\cite{gondolo:sgr1,gondolo:sgr2}, the angle is in the range
$2-15^{\circ}$.  This feature suggests that the presence of a WIMP
stream in the solar neighborhood could be tested for by comparing the
median direction of the observed recoil vectors with the (known)
direction of solar motion.

For $\rho_{\rm str}/\rho_0 = 25\%$ (which is at the upper end of the
range of values suggested in~\cite{gondolo:sgr2,gondolo:sgr1}, and
therefore provides a lower limit on the number of events required), we
find that the minimum number of events required to reject
$(l_{\odot},b_{\odot})$ as the median direction, $N_{\rm dir}$, using
the ${\cal X}^2$ test as defined in Appendix~\ref{dirtest:A} is $201
\, (294)$ for $R_{\rm c}=A_{\rm c}=0.9 \, (0.95)$.  Comparing these
numbers with those required to reject isotropy of recoils for this
stream density, it can be seen that $\sim10-20$ times more events are
needed to reject the direction of the solar motion as the median
recoil direction. These numbers of events would
be achievable with a $10^5 \, {\rm kg \, day}$ exposure for
WIMP-proton cross sections down to $\sigma_0\sim 4\times10^{-8} \,
(6\times10^{-8}) \, {\rm pb}$ for $\rho_{0}= 0.3 \, {\rm GeV \,
cm}^{-3}$.

In the case of axial recoil directions, the ${\cal X}^2$ test cannot
be applied, however, the rotational symmetry test presented in
Appendix~\ref{rotsymtest:A} can be used since the recoil distribution
will not be rotationally symmetric about $(l_{\odot},b_{\odot})$ in
the presence of a stream and we find $N_{\rm rot} = 390 \, (574)$ for
$R_{\rm c}=A_{\rm c}=0.9 \, (0.95)$.  These numbers are roughly twice
those for the ${\cal X}^2$ test, and also for rejecting isotropy with
axial data. Thus with only axial data, a $10^5 \, {\rm kg \, day}$
exposure would be sufficient to identify the presence of a tidal
stream with properties as described in Sec.~\ref{stream} comprising
$25\%$ of the local WIMP density for WIMP-proton cross sections down
to $\sigma_0\sim 8\times10^{-8}\, (1\times10^{-7}) \, {\rm pb}$ and $
\rho_{0} =0.3 \, {\rm GeV \, cm}^{-3}$.

We caution that these numbers have been calculated for parameter
values at the optimistic ends of the ranges estimated in
Ref.~\cite{gondolo:sgr1,gondolo:sgr2}, i.e. high density and low
velocity dispersion. A lower stream density and/or a higher velocity
dispersion would give a peak recoil direction closer to the mean
direction of motion of the Sun, and make the deviation due to the
stream harder to detect. For the most pessimistic case of $\rho_{\rm
str}/\rho_{0} = 0.003$ and $\sigma_{\rm str} = 70 \, {\rm km s^{-1}}$,
the net WIMP flux and recoil rate are virtually indistinguishable from those
for the standard halo model and tens of thousands of events would be
needed to detect the stream. In general the directional detectability
of cold streams of WIMPs will clearly depend on how much their bulk
velocity deviates from the direction of solar motion.

\section{Discussion}
\label{discuss}

We have studied the application of non-parametric tests, developed for
the analysis of spherical
data~\cite{mardia:jupp,fisher:lewis:embleton}, to the analysis of
simulated data as expected from a TPC-based directional WIMP
detector~\cite{drift,sean:drift}, taking into account the
uncertainties in the reconstruction of the nuclear recoil directions.
These tests, unlike likelihood analysis, do not require any
assumptions about the form of the local WIMP velocity
distribution. This is advantageous as even if the properties
(i.e. shape, velocity anisotropy, and density profile) of the MW halo
at the solar radius were accurately determined there would likely be a
wide range of local velocity distributions consistent with these
properties.

For a range of fiducial anisotropic and/or triaxial halo models, with
parameters chosen to reproduce the range of properties found in
simulations/observations of dark matter halos, we calculate the number
of events required to distinguish the WIMP directional signal from an
isotropic background using a variety of tests. The most powerful test
is the, co-ordinate system dependent, test of the mean angle between
the observed nuclear recoils and the direction of motion of the Sun,
which takes advantage of the fact that, for a smooth halo, the recoil
rate is peaked in the direction of the Sun's motion. We find that if
the senses of the recoils are known then of order ten events will be
sufficient to distinguish a WIMP signal from an isotropic background
for all of the halo models considered, with the uncertainties in
reconstructing the recoil direction only mildly increasing the
required number of events. If the senses of the recoils are not known
the number of events required is an order of magnitude larger, with a
large variation between halo models and the recoil resolution is now
an important factor. The number of events required would be
significantly larger if the WIMP velocity distribution {\em in the
rest frame of the detector} is close to isotropic~\cite{copi:krauss},
which could be the case if the MW halo is co-rotating or if the local
dark matter density has a significant contribution from a cold flow
with direction so as to cancel out the front-rear asymmetry from the
smooth background distribution~\cite{sikivie}, however for halos
formed hierarchically (as is the case in Cold Dark Matter cosmologies)
neither of these possibilities are expected to occur (see
e.g. Ref.~\cite{moore:draco,helmi:white:springel}).  Our conclusions
for the case where the recoil senses are known are in broad agreement
with the results of Ref.~\cite{copi:krauss}. It is reassuring that the
realistic modeling of detector properties and the use of statistics
that can be applied to real data do not significantly degrade the
expected detection power of directional experiments.

Halo models which produce significantly different WIMP flux
distributions, unfortunately give very similar recoil rates.
We find that distinguishing between halo models, in particular
determining whether the MW halo is (close to) spherical, using tests
of rotational symmetry will require thousands of events.

Gondolo has shown that the recoil momentum spectrum is the Radon
transform of the WIMP velocity distribution~\cite{gondolo}. In theory
the transform could be inverted to directly measure the WIMP velocity
distribution from the measured recoil momentum distribution, however
the inversion algorithms available in the literature require large
numbers of events~\cite{gondolo}. Furthermore, as we have shown, the
uncertainties in the reconstruction of the recoil directions due to
multiple scattering and diffusion have a significant effect on the
observed recoil distribution and it is not possible to apply this
inversion to real data. Developing techniques to distinguish between
the recoil distributions expected from different halo models, for
finite amounts of data and taking into account the experimental
resolution, is therefore of key importance for the full astronomical
exploitation of data from directional detectors.

If a significant fraction of the local dark matter is in a cold flow
(with velocity dispersion far smaller than its bulk velocity) then the
peak recoil direction will deviate from the direction of the solar
motion.  In fact a stream of halos stars has been found in the solar
neighborhood~\cite{helmi} and using values for the properties of the
stream at the optimistic end of the ranges estimated in
Ref.~\cite{gondolo:sgr1,gondolo:sgr2} this deviation could be detected
with of order a hundred events even if the sense of the nuclear
recoils is not measurable. This illustrates that if the Galactic dark
matter is in the form of WIMPs then WIMP directional detectors, such
as DRIFT~\cite{drift,sean:drift}, will be able to do `WIMP astronomy'.

\begin{acknowledgments}

A.M.G was supported by PPARC and  the Swedish Research Council.
B.M. was supported by PPARC.

We are grateful to John McMillan, Nick Cox, 
Amina Helmi, Vitaly Kudryavtsev and Toby Lewis
for useful discussions. 
\end{acknowledgments}

\appendix

\section{Orbital velocity of the Earth}
 The Earth's orbit around the Sun lies in the Ecliptic
plane and is approximately circular with eccentricity $e=0.0167$ and
semi-major axis $a=1.000 \,$AU$=1.496\times10^{11} \, $m. The tangential and
radial components of the Earth's orbital velocity at a given time $t$ are
calculated from~\cite{green:book}
\begin{eqnarray}
v_{T}(t) & = & \frac{2\pi a(1+e\cos \nu(t))}{P\sqrt{1-e^2}} \,, \nonumber \\
v_{R}(t) & = & \frac{2\pi ae\sin \nu(t)}{P\sqrt{1-e^2}} \,, \nonumber \\
\nu(t) &=& \lambda_{\odot}(t) - \varpi - \pi \,,
\label{earth:velocity}
\end{eqnarray}
where $\lambda_{\odot}(t)$ is the ecliptic longitude of the Sun,
$\varpi=103^{\circ}$~\cite{ast:alm:book} is the argument of perihelion
of the Earth and $P$ is the orbital period. The Solar ecliptic longitude
at time $t$, accurate to $0.01^{\circ}$ for $t$ between 1950 and 2050,
is calculated from~\cite{ast:alm:book} $\lambda_{\odot}(t) = 280.460 +
0.9856474t + 1.915\sin g + 0.020\sin 2g$ where $t$ is the time given
in terms of the Julian Day $JD$ via $t = JD - 2451545.0$ and $g$ is
the mean anomaly of the orbit given by $g = 357.528 + 0.9856003t$.
Using Eqs.~\ref{earth:velocity}, the Earth's orbital velocity in
the Ecliptic coordinate system  is calculated as
\begin{eqnarray}
\vec{v}_{\rm orb}^{\rm E}(t) &=& 
\left( \begin{array}{c}
\mathcal{A}(t)v_{R}(t) -  \mathcal{B}(t)v_{T}(t)\\
\mathcal{A}(t)v_{T}(t) -  \mathcal{B}(t)v_{R}(t)\\
0
\end{array} \right ),\\
\end{eqnarray}
where
\begin{eqnarray}
{\cal A}(t) &=& \cos\nu(t)\cos\varpi - \sin\nu(t)\sin\varpi \,,\\
{\cal B}(t) &=& \sin\nu(t)\cos\varpi - \cos\nu(t)\sin\varpi \,.
\end{eqnarray}
This velocity is then transformed to the galactic coordinate
system via the rotations
\begin{equation}
\vec{v}_{\rm orb}(t) = {\bf G} \, {\bf E} \, \vec{v}_{\rm orb}^{\rm E}(t),
\end{equation}
with rotation matrices
\begin{equation}
{\bf E} = 
\left( \begin{array}{ccc}
1 & 0 & 0 \\
0 & \cos \epsilon & \sin \epsilon \\
0 & -\sin \epsilon & \cos \epsilon
\end{array} \right),
\end{equation}
where $\epsilon=23.5^{\circ}$ is the obliquity of the
ecliptic plane to the equatorial plane of the Earth~\cite{ast:alm:book} and~\cite{green:book} 
\begin{equation}
{\bf G} = 
\left( \begin{array}{ccc}
-0.054876 & -0.873437 & -0.483835\\
0.494109 & -0.444830 & 0.746982\\
-0.867666 & -0.198076 &  0.455984
\end{array} \right).
\end{equation}

\section{Statistical tests}
\label{stat:A}
\subsection{Tests of isotropy}
\label{iso:A}

The simplest coordinate independent statistic for vectorial data is
the so-called modified Rayleigh-Watson statistic ${\cal
W}^{\star}$~\footnote{This modified statistic, and the others
considered in this section, have the advantage of approaching their
large $N$ asymptotic distribution for smaller $N$ than the unmodified
statistic.} . For a sample of $N$ unit vectors $\vec{x}_i $, ${\cal
W}^{\star}$ is defined as~\cite{watson1,watsonbook,mardia:jupp}
\begin{equation}
{\cal W}^{\star} = \left( 1- \frac{1}{2N} \right) {\cal W} +
\frac{1}{10N} {\cal W}^2 \,,
\end{equation}
where ${\cal W}$ is the (unmodified)  Rayleigh-Watson statistic 
\begin{equation}
{\cal W} = \frac{3}{N}  {\cal R}^2 \,,
\end{equation}
and ${\cal R}$ is the Rayleigh statistic:
\begin{equation}
{\cal R} = \left | \sum^N_{i=1} \vec{x}_i \right | \,.
\end{equation}
The value of ${\cal W}^{\star}$
becomes larger as the degree of anisotropy increases and for
isotropically distributed vectors, ${\cal W}^{\star}$ is asymptotically
distributed as $\chi^2_3$~\cite{watson1,watsonbook}.  
We find, using Monte Carlo simulations, that the difference between
$\chi^2_3$ and the true distribution of ${\cal W}^{\star}$ for
isotropic vectors in the large ${\cal W}^{\star}$ tail of the
distribution is less than $2\%$ for $N>30$. For $N<30$ the $\chi^2_3$
distribution significantly underestimates the true probability
distribution and we calculate the probability distribution from 
the exact probability distribution of ${\cal R}$, as described in
Ref.~\cite{stephens:rayleigh}.

The Rayleigh-Watson statistic has the drawback that it is not
sensitive to distributions which are symmetric with respect to the
center of the sphere and therefore can not be used with axial data.
The Bingham statistic ${\cal B}^{\star}$ avoids this problem and is
based on the scatter (or orientation) matrix ${\bf T}$ of the data, 
defined as~\cite{watson2,watsonbook,mardia:jupp}
\begin{equation}
{\bf T} = \frac{1}{N}
\sum^N_{i=1}
\left( \begin{array}{ccc}
x_i x_i & x_i y_i & x_i z_i \\
y_i x_i & y_i y_i & y_i z_i \\
z_i x_i & z_i y_i & z_i z_i
\end{array} \right) \,,
\end{equation}
where $(x_i,y_i,z_i)$ are the components of the $i$-th vector or axis.
This matrix is real and symmetric with unit trace, so that that the
sum of its eigenvalues $e_k$ ($k=1,2,3$) is unity, and for an
isotropic distribution all three eigenvalues should, modulo
statistical fluctuations, be equal to $1/3$. Bingham's modified
statistic ${\cal B}^{\star}$~\cite{bingham,watsonbook,mardia:jupp}
\begin{equation}
{\cal B}^{\star} = {\cal B} \left( 1 - \frac{1}{N}\left[\frac{47}{84} +
\frac{13}{147}{\cal B} + \frac{5}{5292}{\cal B}^2 \right]\right) \,,
\end{equation}
where ${\cal B}$ is the (unmodified) Bingham statistic
\begin{equation}
{\cal B} = \frac{15N}{2}\sum^3_{k=1}\left( e_k - \frac{1}{3}
\right)^2 \,,
\end{equation}
measures the deviation of the eigenvalues $e_k$ from the value
expected for an isotropic distribution. For 
isotropically distributed vectors/axes ${\cal
B}^{\star}$ is asymptotically distributed as $\chi^2_5$, and we have
found, via Monte Carlo simulations, that the difference between the
underlying probability distribution and $\chi^2_5$ in the large ${\cal
B}^{\star}$ tail of the distribution is always less than $10\%$ and is
less than $2\%$ for $N>10$.

The most general test of the uniformity of a sample of $N$ unit
vectors or axes is provided by the statistics of Beran~\cite{beran:stat} and
Gin\'e~\cite{gine:stat}, which are
defined as
\begin{eqnarray}
{\cal A}& = & N - \frac{4}{N\pi} \sum^{N-1}_{i=1} \sum^N_{j=i+1} \psi_{ij}
    \,, \\
{\cal G} & = & \frac{N}{2} - \frac{4}{N\pi} \sum^{N-1}_{i=1} \sum^N_{j=i+1}
\sin\psi_{ij} \,,
\end{eqnarray}
where $\psi_{ij}$ is the angle between the $i$-th and $j$-th
directions.  Beran's statistic ${\cal A}$ tests for distributions
which are asymmetric with respect to the center of the sphere, and
Gin\'e's statistic ${\cal G}$ tests for distributions which are
symmetric with respect to the center of the sphere. A suitable
statistic for testing uniformity against all possible alternatives is
therefore the combination ${\cal F} = {\cal A} + {\cal G}$
~\cite{fisher:lewis:embleton}.  In the case of axial data ${\cal G}$
alone can be used. 
The asymptotic distributions of
${\cal A}$, ${\cal G}$ and ${\cal F}$ have not, to date, been
calculated. We therefore generate the probability
distributions for these statistics under the null hypothesis via Monte
Carlo simulation.

A suitable coordinate system dependent statistic which uses the
fact that, for smooth halo models, the WIMP recoil distribution is
expected to be peaked about $(l_{\odot},b_{\odot})$is~\cite{briggs} 
\begin{equation} 
\langle \cos \theta \rangle = \frac{\sum_{i=1}^N \cos
\theta_i}{N} \,, 
\end{equation} 
for vectorial data, 
\begin{equation}
\langle |\cos \theta| \rangle = \frac{\sum_{i=1}^N |\cos \theta_i|}{N}
\,, 
\end{equation} 
for axial data, where $\theta_i$ is the angle
between $(l_{\odot},b_{\odot})$ and the $i$th vector/axis.  For $N$
isotropic vectors $\langle \cos \theta \rangle \, (\langle |\cos
\theta| \rangle$) can take values on the interval $[-1,1] \, ([0, 1])$
and, due to the central limit theorem, has a gaussian distribution
with mean $0 \, (0.5)$ and variance $1/3N \, (1/3N)$~\cite{briggs}.
The larger the concentration of recoil directions towards
$(l_{\odot},b_{\odot})$ the larger these statistics will be.
We calculate the probability distribution function of $ \langle
\cos \theta \rangle$ and $\langle |\cos \theta| \rangle$ as a function
of $N$ for the null hypothesis of an isotropic recoil spectrum using
Monte Carlo simulations.

\subsection{Tests for rotational symmetry}
\label{rotsymtest:A}

A test of rotational symmetry  about some
hypothesized direction $(\theta_0,\phi_0)$ (valid for vectors or axes) 
can be performed by first
rotating the sample vectors or axes so that their polar angles are
measured relative to $(\theta_0, \phi_0)$. The resultant azimuthal angles 
are then divided by $2 \pi$ ($X_{i}=\phi_{i}/ 2 \pi$) 
and sorted in ascending order.
For rotational symmetry, the distribution of the ordered normalized angles
should be uniform between 0 and 1. This hypothesis can be tested by
using the Kuiper statistic~\cite{fisher:lewis:embleton}, which is 
related to the well-known Kolmogorov-Smirnov test, but 
has the advantage of being invariant
under cyclic transformations and equally
sensitive to deviations at all values of $X$.
The modified Kuiper statistic is defined as 
\begin{equation}
{\cal V}^{\star} = {\cal V} \left(N^{1/2} +0.155+ 
     \frac{0.24}{N^{1/2}}\right) \,,
\end{equation}
where ${\cal V}$ is the (unmodified) Kuiper statistic 
\begin{equation}
{\cal V} = {\cal D}^{+} + {\cal D}^{-} \,,
\end{equation}
and
\begin{eqnarray}
{\cal D}^{+}  &=& {\rm max} \left( \frac{i}{N} - X_i \right),\;\;\;
i=1,\ldots,N\\
{\cal D}^{-} &=& {\rm max}\left( X_i - \frac{i-1}{N} \right) \,.
\end{eqnarray}
As there is no general formula for the distribution of ${\cal
V}^{\star}$ under the null hypothesis of rotational symmetry, we use
Monte Carlos to generate the null probability distribution assuming
the standard halo model~\footnote{ This may appear to introduce a
model dependence into the test, however any other distribution
rotationally symmetric about
$(l_{{\odot}},b_{{\odot}})$ would give the
same null distribution for $V^{\star}$. }.

\subsection{Tests for mean direction}
\label{dirtest:A}

For vectorial data, the median direction is defined as
the direction $(\hat{\theta},\hat{\phi})$ which minimizes
the sum of arclengths between $(\hat{\theta},\hat{\phi})$ and the sample
vectors~\cite{fisher:lewis:embleton}, and can found
by minimizing the quantity
\begin{equation}
{\cal M} = \sum^N_{i=1} \cos^{-1}(\hat{x}x_i + \hat{y}y_i + \hat{z}z_i) \,,
\end{equation}
while for axial data an estimate of the principal axis of the
distribution is provided by the eigenvector corresponding to the
largest eigenvalue of the scatter matrix ${\bf T}$.  A test of
compatibility between the measured median direction and a hypothesized
median direction $(\theta_0,\phi_0)$ for vectorial data can be
performed~\cite{fisher:lewis:embleton} by first rotating the data so
that the polar angles are measured relative to a pole in the direction
of the sample median $(\hat{\theta},\hat{\phi})$ and then calculating
the matrix ${\bf \Sigma}$
\begin{equation}
{\bf \Sigma} = \frac{1}{2} \left( \begin{array}{cc} 
\sigma_{11} & \sigma_{12}\\ \sigma_{21} & \sigma_{22}\\ \end{array} \right) \,,
\end{equation}
where
\begin{eqnarray}
\sigma_{11} &=& 1 + \frac{1}{N}\sum^{N}_{i=1}\cos 2\phi^{'}_i \,, \\
\sigma_{22} &=& 1 - \frac{1}{N}\sum^{N}_{i=1}\cos 2\phi^{'}_i \,, \\
\sigma_{12} &=& \sigma_{21} = \frac{1}{N}\sum^{N}_{i=1}\sin 2\phi^{'}_i  \,,
\end{eqnarray}
from the rotated sample vectors. The sample
vectors are rotated again so that their polar angles are measured
relative to a pole in the direction of the hypothesized median using
and the vector
\begin{equation}
\vec{U} = N^{-1/2} \left(\begin{array}{cc} 
 \sum \cos\phi^0_i\\\sum\sin\phi^0_i\\ \end{array} \right) \,,
\end{equation}
where $\phi^0_i$ is the azimuthal angle of the ith sample
vector relative to a pole at $(\theta_0,\phi_0)$ is calculated.
Finally the test statistic ${\cal X}^{2}$, is defined as
\begin{equation}
{\cal X}^2 = \vec{U}^T \mathbf{\Sigma}^{-1}\vec{U} \,,
\end{equation}
and is distributed as
$\chi^2_2$ for samples drawn from a distribution with median direction
$(\theta_0,\phi_0)$.

\section{Hypothesis testing}
\label{hyptest}

For each halo model considered we use Monte Carlo simulations to
generate $N$ recoil scattering events in each of $10^{5}$ experiments,
for values of $N$ between 5 and 400 (the lower value corresponding to
the point at which an anomalous recoil signal would first be
identified at high confidence).   For each experiment the test
statistic ${\cal T}$ is calculated from the $N$ recoil directions used
to give the probability distribution of the statistic, $p_1({\cal
T};N)$, as a function of the number of recoil events.  The probability
distribution for the null hypothesis of isotropy, $p_0({\cal T};N)$,
is calculated using analytical expression where available and
otherwise via Monte Carlo simulations. As in standard hypothesis
testing, the overlap between these two distributions allows the
probability~\footnote{We use the frequentist definition of
`probability' throughout.} with which the null and alternative
hypotheses can be rejected or accepted to be calculated. For a given
value ${\cal T}_{\rm g}$, the rejection factor $R$ is the probability
of measuring ${\cal T} \leq {\cal T}_{\rm g}$ if the null hypothesis
is true:
\begin{equation}
 R = \int_0^{{\cal T}_{\rm g}} p_0({\cal T}; N) \,{\rm d} {\cal T} \,.
\end{equation}
The rejection factor thus gives the confidence level with which the
null hypothesis can be rejected given a particular value of the test
statistic ${\cal T}={\cal T}_{\rm g}$. For the same value of the test
statistic ${\cal T}_{\rm g}$, the acceptance factor $A$ is the
probability of measuring ${\cal T} \geq {\cal T}_{\rm g}$ if the
alternative hypothesis is true
\begin{equation}
A = \int_{{\cal T}_{\rm g}}^{\infty} p_1({\cal T}; N) \, {\rm d} {\cal T} \,.
\end{equation}
Equivalently, under the frequentist
definition of probability, it is the fraction of experiments in which
the alternative hypothesis is true that measure
${\cal T} \geq {\cal T}_{\rm g}$ and thus reject the null hypothesis at
confidence level $R$.

By calculating $R$ and $A$ from the probability distributions of the
test statistic for the null and alternative hypothesis as a function
of $N$, an `acceptance-rejection' plot can be built up for each value
of $N$ and for any given level of rejection, $R_{\rm c}$, the level of
acceptance $A_{\rm c}$ achievable for each recoil sample size $N$
calculated.  In other words we find the $N$ such that `for $N$
observed recoils, the null hypothesis is rejected at the $100 R_{\rm
c}\%$ confidence level in $100 A_{\rm c}\%$ of experiments in which
the alternative hypothesis is true'.

Clearly, a high value of $R_{\rm c}$ is required to reject the null
hypothesis at high confidence. A high acceptance is also required; if,
for instance, $A_{\rm c}=0.1$ then only 1 in 10 experiments will be
able to reject the null hypothesis at the given $R_{\rm c}$,
furthermore if $A_{\rm c}$ is low, the null hypothesis could sometimes
be erroneously rejected at a given confidence level with a low number
of events due to statistical fluctuations.  We therefore use $A_{\rm
c}=R_{\rm c}=0.9 \, (0.95)$ as our criteria, and calculate the
corresponding minimum number of events required.



\begin{thebibliography}{}

\bibitem{amtheory} A. K. Drukier, K. Freese and D. N. Spergel, Phys. Rev. 
            D {\bf 33}, 3495 (1986); K. Freese, J. Frieman and A. Gould,
             Phys. Rev. D {\bf 37}, 3388 (1988).
\bibitem{dirndep} D. N. Spergel, Phys. Rev. 
            D {\bf 37}, 1353 (1988).
\bibitem{jkg} G. Jungman, M. Kamionkowski and K. Griest, Phys. Rep. {\bf 267}, 
               195 (1996).
\bibitem{ls} J. D. Lewin and P. F. Smith, Astropart. Phys. {\bf 6}, 87
             (1996).
\bibitem{dama}   R. Bernabei et al., Phys. Lett. {\bf B389}, 757 (1996); 
               ibid {\bf B408}, 439 (1997); ibid {\bf B424}, 195 (1998); 
                 ibid {\bf B450}, 448 (1999); ibid {\bf B480}, 23 (2000);
                 Riv. Nuovo. Cim. {\bf 26N1} 1 (2003), {\tt astro-ph/0307403}.


\bibitem{otherexpt}  D. Akerib et al.,  {\tt astro-ph/0405033};  
                    A. Benoit et al., Phys. Lett. {\bf B545}, 43 (2002);
                  N. J. Smith et al., proceedings of
               4th Int. Workshop on Identification of 
                Dark Matter (York, 2002) ed. N. J. C. Spooner and 
                 V. Kudryavtsev, World Scientific,
                 Singapore (2003). 
\bibitem{amhalo}  M. Brhlik and L. Roszkowski, 
                  Phys. Lett. {\bf B464}, 303 (1999), {\tt astro-ph/9903468};
                  G. Gelmini and P. Gondolo, {\tt hep-ph/0405278}.   
\bibitem{amevans}  P. Belli et al., Phys. Rev. D
              {\bf 61}, 023512 (2000), {\tt hep-ph/0203242};
               C. J. Copi and L. M. Krauss,  Phys. Rev. D {\bf 67}, 103507,
                 (2003) {\tt  astro-ph/0208010},
                   N. Fornengo and S. Scopel, Phys. Lett. B {\bf 576}, 
                 189 (2003), {\tt hep-ph/0301132}.
\bibitem{amhalog1}  A. M. Green, Phys. Rev. D {\bf 63},
             043005 (2001), {\tt astro-ph/0008318}.
\bibitem{amhalog2}    A. M. Green, Phys. Rev. D   {\bf 68}, 
               023004 (2003), {\tt astro-ph/0304446}.
\bibitem{spindep}  P. Ullio, M. Kamionkowski and P. Vogel, 
                 JHEP {\bf 0107}, 044 (2001),
                         {\tt hep-ph/0010036}, 
          A. Kurylov and M. Kamionkowski, Phys. Rev. D {\bf 69},
                            063503 (2004), {\tt hep-ph/0307185}.

\bibitem{inelas} D. Smith and N. Weiner, Phys. Rev. D {\bf 64}, 
                     043502 (2001), {\tt hep-ph/0101138};
                        D. Smith and N. Weiner, {\tt hep-ph/0402065}.
\bibitem{dirrot} B. Morgan et al.,  Proccedings of The International
                Workshop on Technique and Application of Xenon Detectors,
                Tokyo, Japan 2000, p78
             eds. Y. Suzuki, M. Nakahata, Y. Koshio and 
              S. Moriyama, World Scientific (2002); B.Morgan,
             Nucl. Inst. and Meth. A {\bf 513}, 226 (2003).
\bibitem{lhedet} S. R. Bandler et al., Phys. Rev. Lett. {\bf 74}, 3169 (1995).
\bibitem{bafdet} R. J. Gaitskell et al., Nucl. Inst. and Meth. A
            {\bf 370}, (1996).

\bibitem{stildet} H. Sekiya et al., Phys. Lett. B {\bf 571}, 132 (2003);
                  H. Sekiya et al., to appear in proccedings of
                  5th workshop on Neutrino Oscillations and their Origian
                  (NOON2004), {\tt astro-ph/0405598}.
\bibitem{drift}  D. P. Snowden-Ifft, C. J. Martoff, and J. M. Burwell, 
                 Phys. Rev. D  {\bf 61}, 1 (2000), 
                   {\tt astro-ph/9904064}.
\bibitem{sean:drift} G. J. Alner et al., to appear in Nucl. Inst. and Meth. A.
\bibitem{newage} T. Tanimori et al., Phys. Lett. B {\bf 578}, 241 (2004);
                 {\tt astro-ph/0310638}.
\bibitem{copi:krauss} C. J. Copi, J. Heo and L. M. Krauss, 
                       Phys. Lett. B {\bf 461}, 43 (1999),
                                      {\tt astro-ph/9904499};
                                     C. J. Copi and L. M. Krauss, 
                       Phys. Rev. D {\bf 63}, 043507 (2001), 
                                    {\tt astro-ph/0009467}.       
\bibitem{lehner:dir}  M. J. Lehner et al.,  
                     {\em Dark Matter in Astro and Particle Physics},
                                  Proceedings of the International 
                        Conference DARK2000, Heidelberg, 
                                   Germany, 2000, p590 ed. 
                       H. V.  Klapdor-Kleingrothaus, Springer-Verlag (2001). 

\bibitem{shape} C. S. Frenk, S. D. M. White, M. Davis and G. Estafthiou, 
              Astrophys. J. {\bf 237}, 507 (1988).
\bibitem{js} Y. P. Jing and Y. Suto, Astrophys. J. {\bf 574}, 538 (2002), 
             {\tt astro-ph/0202064}.
\bibitem{moore:draco} B. Moore et al., Phys. Rev. D {\bf 64}, 063508 (2001), 
                     {\tt astro-ph/0106271}.
\bibitem{sph} J. Dubinski and R. G. Carlberg, Astrophys. J. {\bf 378}, 
            496 (1991);  S. Kazantzidis et al., 
            Astrophys. J. {\bf 611}, L63 (2004). {\tt astro-ph/0405189}.
\bibitem{shapeobs}   P. D. Sackett,  Galaxy Dynamics, ASP Conf Series, 182, 
               p393, eds. D. Merritt, J.A. Sellwood and M. Valluri, (1999),
                       {\tt astro-ph/9903420};
                  M. R. Merrifield,  to appear in proceedings of
           IAU Symposium 220 `Dark Matter in Galaxies', Sydney,
         2003  eds. D. J. Pisano, M. Walker and K. Freeman, Astron.
        Soc. Pacific 
            (2004) {\tt astro-ph/0310497}.
\bibitem{logellip} J. J. Binney,  Mon. Not. Roy. Astron. Soc. {\bf
               196}, 455 (1981); P. T. de Zeeuw, and D, Pfenniger, 
          Mon. Not. Roy. Astron. Soc. {\bf
               235}, 949 (1988) Erratum: {\bf
               262}, 1088.
\bibitem{evans} N. W. Evans, C. M. Carollo and P. T. de Zeeuw, 
                  Mon. Not. Roy. Astron. Soc. {\bf 318},
                        1131, (2000), {\tt astro-ph/0008156}.
\bibitem{osipkov} L. P. Osipkov, Pis'ma Astron, Zh. {\bf 55}, 77 (1979).
\bibitem{merritt} D. Merritt, Astron. J. {\bf 90}, 1027 (1985).
\bibitem{ullio:kamionkowski}  P. Ullio and M. Kamionkowski, 
                JHEP {\bf 0103},  049 (2001), {\tt hep-ph/0006183}.
\bibitem{evansomgreen}  A. M. Green, Phys. Rev. D   {\bf 66}, 
                     083003 (2002), {\tt astro-ph/0207336}.
\bibitem{NFW}  J. F. Navarro, C. S. Frenk and S. D. M. White, 
                    Astrophys. J. {\bf 462}, 563 (1996).
\bibitem{widrow:fitting} L. M. Widrow, Astrophys. J. Suppl. S. 
             {\bf 131}, 39 (2000), {\tt astro-ph/0003302}.
\bibitem{stiff:widrow:frieman}  D. Stiff, L. M. Widrow and J. Frieman,  
                       Phys. Rev. D {\bf 64},
                             083516 (2001), {\tt astro-ph/0106048}.
\bibitem{helmi:white:springel} A. Helmi, S. D. M. White and V. Springel, 
                     Phys. Rev. D {\bf 66},
                  063502 (2002), {\tt astro-ph/0201289}.
\bibitem{stiff:widrow2}    D. Stiff and  L. M. Widrow, 
                    Phys. Rev. Lett. {\bf 90}, 211301 (2003),
                           {\tt astro-ph/0301301}.                             
\bibitem{gondolo:sgr1}  K. Freese, P. Gondolo and H. J. Newberg, 
                          {\tt astro-ph/0309279}.
\bibitem{mass:sgr} S. R. Majewski et al., Astrophys. J. {\bf 599}, 
                          1082 (2003), {\tt astro-ph/0304198}. 
\bibitem{sdss:sgr} H. J. Newberg et al., Astrophys. J. {\bf 596}, 
                     L191 (2003), {\tt astro-ph/0309162}.
\bibitem{helmi} A. Helmi, S. D. M. White, T. P. de Zeeuw and H. Zhao,
                Nature, {\bf 402}, 53 (1999), {\tt astro-ph/9911041}.
\bibitem{gondolo:sgr2}   K. Freese, P. Gondolo, H. J. Newberg and M. Lewis, 
                         Phys. Rev. Lett. {\bf 92},
                               111301 (2004), {\tt astro-ph/0310334}. 
\bibitem{helmi2} A. Helmi, talk at 5th International Workshop on the 
          Identification of Dark Matter, Edinburgh 2004,
          http://www.shef.ac.uk/physics/idm2004/talks/
\bibitem{bellazini}R. Bellazzini and G. Spandre, Nucl. Inst. and
  Meth. A. {\bf 513}, 231, (2003).
\bibitem{ohnuki} T. Ohnuki, D. P. Snowden-Ifft and C. J. Martoff,
     Nucl. Inst. and Meth. A. {\bf 463}, 142, (2001).
\bibitem{srim} J. F. Ziegler, J. P. Biersack and U. Littmark, 
            {\em The stopping and range
          of ions in solids}, Pergamon Press (1985), http://www.srim.org.

\bibitem{quench} D. P. Snowden-Ifft, T. Ohnuki, E. S. Rykoff 
             and C. J. Martoff, Nucl. Inst. and Meth. A. 
          {\bf 498}, 155, (2003).

\bibitem{binney:tremaine} J. Binney and S. Tremaine, 
                {\em Galactic Dynamics}, Princeton University Press (1987).

\bibitem{dehnen:binney:hipparcos} W. Dehnen and J. J. Binney, 
                             Mon. Not. Roy. Astron. Soc. {\bf 298},
                                       387 (1998), 
                                  {\tt astro-ph/9710077}.

 \bibitem{mardia:jupp} K. V. Mardia and P. Jupp, 
                     {\em Directional Statistics},  Wiley, Chichester (2002).
\bibitem{fisher:lewis:embleton}  N. I. Fisher, T. Lewis and B. J. J. Embleton,
         {\em Statistical analysis of spherical data}, CUP, (1987).

\bibitem{drift2:design}  M. J. Carson et al. to appear in 
            Nucl. Inst. and Meth. A. {\tt hep-ex/0503017};
                   J. C. Davies talk at 5th International Workshop on the 
          Identification of Dark Matter, Edinburgh 2004,
          http://www.shef.ac.uk/physics/idm2004/talks/,
          M. J. Carson et al., Astropart. Phys. {\bf 21}, 667 (2004).

\bibitem{sikivie} P. Sikivie, I. I. Tkachev and Y. Wang, 
              Phys. Rev. Lett. 
            {\bf 75}, 2911 (1995), Phys. Rev. D {\bf 56}, 1863 (1997); 
             P. Sikivie, Phys. Lett. B {\bf 432}, 139 (1998). 
\bibitem{gondolo} P.Gondolo, Phys. Rev. D {\bf 66}, 103513 (2002), 
               {\tt hep-ph/0209110}.
\bibitem{green:book} R. M. Green, {\em Spherical Astronomy}, CUP (1993).
\bibitem{ast:alm:book}  {\em The Astronomical Almanac for the year 2003}, 
                   United States Government Printing Office (2003).
\bibitem{watson1}  G. S. Watson, Geophys. Suppl. Mon. Not. Roy. 
                     Astron. Soc. {\bf 7}, 160 (1956).
\bibitem{watsonbook}   G. S. Watson, {\em Statistics on Spheres}, 
                      Wiley, New York (1983).
\bibitem{stephens:rayleigh}  M. A. Stephens, J. Amer. Statist. 
                       Assoc.  {\bf 59}, 160 (1964).
\bibitem{watson2} G. S. Watson,  J. Geol. {\bf 74}, 786 (1966).
\bibitem{bingham}  C. Bingham, Ann. Stat. {\bf 2}, 1201 (1974).
\bibitem{beran:stat} R. Beran, J. App. Prob. {\bf 5}, 177 (1968).
\bibitem{gine:stat} E. M. Gin\'e, Ann. Stat. {\bf 3}, 1243 (1975). 
\bibitem{briggs} M. S. Briggs,  Astrophys. J. {\bf 407}, 125, (1993).    

\end{thebibliography}
\end{document}